\documentclass[twoside,twocolumn,aps,pra,superscriptaddress,showpacs]{revtex4-1}
\usepackage[T1]{fontenc}
\usepackage[latin9]{inputenc}
\usepackage{color}
\usepackage{amsmath}
\usepackage{amssymb}
\usepackage{graphicx}
\usepackage{esint}
\usepackage[unicode=true,pdfusetitle,
 bookmarks=true,bookmarksnumbered=false,bookmarksopen=false,
 breaklinks=false,pdfborder={0 0 1},backref=false,colorlinks=false]
 {hyperref}

\makeatletter

\providecommand{\tabularnewline}{\\}

\newcommand{\rb}{\rangle\kern-0.3em\rangle}
\newcommand{\lb}{\langle\kern-0.3em\langle}

\newcommand{\Tr}{\mathop{\mathrm{Tr}}}

\newcommand{\beq}{\begin{equation}}
\newcommand{\eeq}{\end{equation}}
\newcommand{\beqn}{\begin{eqnarray}}
\newcommand{\eeqn}{\end{eqnarray}}

\newcommand{\hc}{\mathrm{h.c.}}
\newcommand{\1}{\leavevmode{\rm 1\ifmmode\mkern  -4.8mu\else\kern -.3em\fi I}}

\usepackage{times}

\makeatother

\begin{document}

\title{Topological Protection of Coherence in a Dissipative Environment}

\author{Lorenzo Campos Venuti}

\affiliation{Department of Physics and Astronomy, University of Southern California,
CA 90089, USA}

\author{Zhengzhi Ma}

\affiliation{Department of Physics and Astronomy, University of Southern California,
CA 90089, USA}

\author{\textcolor{black}{Hubert Saleur}}

\affiliation{Department of Physics and Astronomy, University of Southern California,
CA 90089, USA}

\affiliation{Institut de Physique Théorique CEA Saclay 91191 Gif Sur Yvette Cedex
France}

\author{Stephan Haas}

\affiliation{Department of Physics and Astronomy, University of Southern California,
CA 90089, USA}

\date{\today}
\begin{abstract}
One dimensional topological insulators are characterized by edge states
with exponentially small energies. According to one generalization
of topological phase to non-Hermitian systems, a finite system in
a non-trivial topological phase displays surface states with exponentially
long life times. In this work we explore the possibility of exploiting
such non-Hermitian topological phases to enhance the quantum coherence
of a fiducial qubit embedded in a dissipative environment. We first
show that a network of qubits interacting with lossy cavities can
be represented, in a suitable super-one-particle sector, by a non-Hermitian
``Hamiltonian'' of the desired form. We then study, both analytically
and numerically, one-dimensional geometries with up to three sites
per unit cell, and up to a topological winding number $W=2$. For
finite-size systems the number of edge modes is a complicated function
of $W$ and the system size $N$. However we find that there are precisely
$W$ modes localized at one end of the chain. In such topological
phases the quibt's coherence lifetime is exponentially large in the
system size. We verify that, for $W>1$, at large times, the Lindbladian
evolution is approximately a non-trivial unitary. For $W=2$ this
results in Rabi-like oscillations of the qubit's coherence measure. 
\end{abstract}

\maketitle

\section{Introduction}

There is a growing interest in the study of non-Hermitian generalizations
of topological phases of matter \cite{rudner_topological_2009,rudner_phase_2010,esaki_edge_2011,diehl_topology_2011,bardyn_topology_2013,zeuner_observation_2015,malzard_topologically_2015,mogilevtsev_quantum_2015,rudner_survival_2016,lee_anomalous_2016}
which can be observed in dissipative systems. Topological features
are potentially useful, as they tend to be robust with respect to
small perturbations and local noise sources. In this work we explore
the possibility of exploiting such, non-trivial, non-Hermitian topological
phases to protect the coherence of a preferential qubit in a network
of dissipative cavities.

Since eigenvalues of non-Hermitian matrices are complex there are
at least two possible definitions of topological phases in non-Hermitian
systems \cite{esaki_edge_2011,rudner_survival_2016}. These definitions
differ in how one generalizes the Hermitian notion of gap: namely
one can consider either the real or the imaginary part of the eigenvalues.
According to the imaginary-part classification of Ref.~\cite{rudner_survival_2016},
as a consequence of a generalized \emph{bulk-edge} correspondence,
a non-trivial topological dissipative phase is characterized at finite
size by the presence of quasi-dark states localized at the boundary
of the system. By quasi-dark states we mean eigenstates of the system
that have a decay time exponentially large in the system size. It
is natural to expect that this feature may be useful to protect quantum
coherence. Indeed, as we will show, if a fiducial qubit is placed
at one end of a linear system, both these features, localization and
\emph{dark}ness, conspire to preserve its coherence in a well defined
way.

In recent experiments such non-Hermitian systems \textendash{} in
fact essentially non-Hermitian quantum walks \textendash{} can be
observed in classical waveguides using the analogy between Helmoltz
and Schrödinger equation \cite{zeuner_observation_2015}. In Ref.~\cite{rudner_topological_2009}
it was proposed that a non-Hermitian version of the Su-Schrieffer-Heeger
(SSH) model \cite{su_solitons_1979} could emerge from a single resonator
described by a Jaynes-Cummings model in the semi-classical, large-photon
number regime.

Here we consider a network of dissipative cavity resonators interacting
á-la Jaynes-Cummings. This model is known to describe the physics
of many experimental quantum platforms, ranging from superconducting
qubits to arrays of microcavities \cite{jarlov_effect_2016}. We show
that, in an appropriate \emph{super}-one-particle sector, the Lindbladian
is precisely given by a non-Hermitian quantum walk determined by the
network geometry. Moreover, the coherence of a preferential qubit
in the network is \emph{exactly} described by the Schrödinger evolution
with such a ``non-Hermitian Hamiltonian''.

Having in mind the goal of prolonging the coherence, we analyze analytically,
and confirm numerically, the behavior of the coherence for various
finite size networks. The simplest of such a networks is a non-Hermitian
tight-binding model with a single, both diagonal and off-diagonal,
impurity. We then consider \emph{topologically non-trivial models},
such as a non-Hermitian SSH model, that can have topological charge
zero or one. In finite size, there are always two dark modes for $N$
odd while there is one quasi-dark mode in the topologically non-trivial
sector for $N$ even. However there is always (irrespective of $N$)
a dark or quasi-dark mode localized at one end of the chain. An analogous
situation is found in models with three sites per unit cell, were
the topological winding number $W$ can be zero, one or two. The exact
number of quasi-dark modes is not a simple function of $W$ alone.
However we find precisely $W$ dark or quasi-dark modes localized
at one end of the chain. In the case $W=2$, the long-time dynamics
of the dissipative network becomes unitary, spanning a two-dimensional
space were the coherence shows Rabi-like oscillations.

\begin{figure}[b]
\centering{}\includegraphics[width=0.45\textwidth]{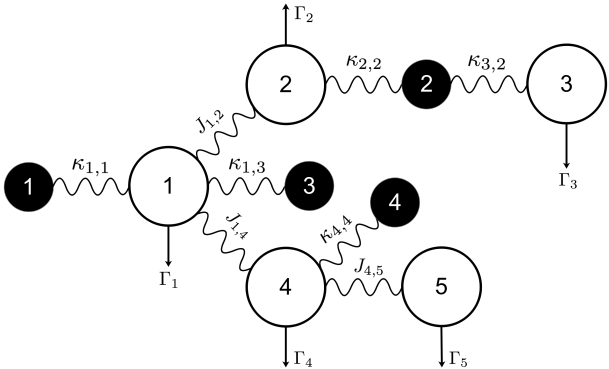} \caption{A general network of qubits interacting with lossy cavities. Wavy
lines indicate coherent hopping and straight arrows incoherent decay.
White dots represent (leaky) cavities while black dots are (long-lived)
two-level systems (qubit). \label{fig:network}}
\end{figure}

\section{Setting the stage}

Our model is a network of dissipative cavities (modes) interacting
with two-level systems (qubit) in a Jaynes-Cummings fashion. To make
it more general, we allow qubits to interact with more than one cavity,
although this may be experimentally challenging to realize. We imagine
a network of $M$ qubits interacting with $K$ cavity modes. Excitations
can hop from mode to mode and also from qubit to mode. At this stage
we don't include hopping from qubit to qubit, as this is definitely
harder to realize. Our goal will be to monitor, and possibly enhance,
the coherence of a fiducial qubit in this network.

We assume the standard rotating-wave approximation, such that the
coherent part of the evolution is given by the following Hamiltonian:
\begin{align}
H & =\sum_{i=1}^{M}\omega_{i}^{0}\sigma_{i}^{z}+\sum_{l,m=1}^{K}J_{l,m}(a_{l}^{\dagger}a_{m}+\hc)\\
 & +\sum_{l=1}^{K}\omega_{l}a_{l}^{\dagger}a_{l}+\sum_{i=1}^{M}\sum_{l=1}^{K}\kappa_{l,i}(a_{l}^{\dagger}\sigma_{i}^{-}+\hc),
\end{align}

where $a_{l}^{\dagger}$ and $a_{l}$ are the creation and annihilation
operators for the cavity mode $l$ and $\sigma_{i}^{\pm}$ are the
ladder operators for qubit $i$. On top of this, cavities leak photons
at rate $\Gamma_{l}$. A Lindblad master equation for the system can
be written as $\dot{\rho}=\mathcal{L}[\rho]$ with $\mathcal{L}=\mathcal{K}+\mathcal{D}$.
The coherent term is $\mathcal{K}=-i\left[H,\bullet\right]$ and the
dissipative part reads 
\begin{equation}
\mathcal{D}[\rho]=\sum_{l=1}^{K}\Gamma_{l}[a_{l}\rho a_{l}^{\dagger}-\frac{1}{2}\{a_{l}^{\dagger}a_{l},\rho\}],\label{lindbladian}
\end{equation}

i.e., we assume sufficiently low temperatures such that no photons
are excited via interaction with the bath. This form of the dissipation
is consistent with the cavity physics whereby essentially only the
cavity modes decay whereas the two level-systems (corresponding to
some hyperfine level of an atom in the cavity) are extremely long-lived
and decay only indirectly through interaction with the cavity. An
example of such a dissipative network with $M=4$ and $K=5$ is schematically
depicted in Fig.~\ref{fig:network}. Let $i=1$ indicate the fiducial
qubit. In order to study the evolution of the qubit's coherence, we
initialize it in a pure state $\alpha_{0}|\uparrow\rangle+\beta_{0}|\downarrow\rangle$,
while we require that all cavities be empty and all other qubits in
the $|\downarrow\rangle$ state. We denote with $|0\rangle$ the overall
vacuum (cavities with no photons and qubits in the $|\downarrow\rangle$
state) and $|j\rangle$, $j=1,\ldots,N\equiv M+K$ the state with
an excitation, either bosonic or spin-like, at position $j$, with
$j=1$ denoting the fiducial qubit and $j=2,3,\ldots,N$ the remaining
cavities/qubits. With this initial condition the relevant Hilbert
space is $\mathcal{H}=\mathrm{Span}\left\{ |0\rangle,|j\rangle,j=1,\ldots,N\right\} $,
and the dynamics are restricted to the space $\mathcal{V}=L(\mathcal{H})$.
A density matrix in $\mathcal{V}$ has the form 
\begin{align}
\rho & =\rho_{0,0}|0\rangle\langle0|+\left(\sum_{j=1}^{N}\rho_{0,j}|0\rangle\langle j|+\hc\right)\\
 & +\sum_{i,j=1}^{N}\rho_{i,j}|i\rangle\langle j|.
\end{align}
After tracing out all but the fiducial qubit degrees of freedom, the
reduced qubit density matrix reads 
\begin{align}
\rho^{\mathrm{qubit}} & =\left(\rho_{0,0}+\sum_{i=2}^{N}\rho_{i,i}\right)|\downarrow\rangle\langle\downarrow|\nonumber \\
 & +\left(\rho_{0,1}|\downarrow\rangle\langle\uparrow|+\hc\right)+\rho_{1,1}|\uparrow\rangle\langle\uparrow|.\label{eq:rho_qubit}
\end{align}

A coherence measure of the qubit can be defined as \cite{baumgratz_quantifying_2014}
\begin{equation}
\mathcal{C}(t)=\sum_{i,j(i\neq j)}\left|\rho_{i,j}^{\mathrm{qubit}}(t)\right|.\label{eqn: coherence}
\end{equation}
Using equation \eqref{eq:rho_qubit} we obtain $\mathcal{C}=2\left|\rho_{0,1}\right|$.

\section{Mapping to a non-Hermitian tight-binding model\label{sec:Mapping-to-a}}

If we initialize the system with at most one excitation, the Lindbladian
generates states with at most one excitation and the dynamics are
contained in the sector $\mathcal{V}$. We are then led to consider
the following linear spaces $\mathcal{V}_{0,0}=\mathrm{Span}\left(|0\rangle\langle0|\right)$,
$\mathcal{V}_{0,1}=\mathrm{Span}\left(\left\{ |0\rangle\langle j|,j=1,\ldots,N\right\} \right)$,
$\mathcal{V}_{1,0}=\mathrm{Span}\left(\left\{ |j\rangle\langle0|,j=1,\ldots,N\right\} \right)$
and $\mathcal{V}_{1,1}=\mathrm{Span}\left(\left\{ |i\rangle\langle j|,\,i,j=1,\ldots,N\right\} \right)$.
The Hamiltonian conserves the number of excitations so the coherent
part $\mathcal{K}$ is block diagonal in the reduced space $\mathcal{V}=\mathcal{V}_{0,0}\oplus\mathcal{V}_{0,1}\oplus\mathcal{V}_{1,0}\oplus\mathcal{V}_{1,1}$.
Moreover 
\begin{align}
\mathcal{D}(|0\rangle\langle0|) & =0\\
\mathcal{D}(|0\rangle\langle j|) & =-\frac{\Gamma_{j}}{2}|0\rangle\langle j|\\
\mathcal{D}(|i\rangle\langle j|) & =\Gamma_{i}\delta_{i,j}|0\rangle\langle0|-\frac{1}{2}\left(\Gamma_{i}+\Gamma_{j}\right)|i\rangle\langle j|.
\end{align}
Note that $\Gamma_{i}=0$ for $i=$ qubit site, as we are ignoring
the spontaneous decay of the qubits (typically much smaller than cavity
loss rate). This implies that on $\mathcal{V}$ the Lindbladian has
the following block-structure (asterisks denote the only non-zero
elements) in $\mathcal{V}=\mathcal{V}_{0,0}\oplus\mathcal{V}_{0,1}\oplus\mathcal{V}_{1,0}\oplus\mathcal{V}_{1,1}$
\begin{equation}
\left.\mathcal{L}\right|_{\mathcal{V}}=\left(\begin{array}{cccccccc}
0 &  &  &  &  & * & * & *\\
 & * & *\\
 & * & *\\
 &  &  & * & *\\
 &  &  & * & *\\
 &  &  &  &  & * & * & *\\
 &  &  &  &  & * & * & *\\
 &  &  &  &  & * & * & *
\end{array}\right).
\end{equation}
We also call $\tilde{\mathcal{L}}=\left.\mathcal{L}\right|_{\mathcal{V}_{0,1}}$
the restriction of $\mathcal{L}$ to $\mathcal{V}_{0,1}$ and, in
this basis, one has $\left.\mathcal{L}\right|_{\mathcal{V}_{1,0}}=\overline{\tilde{\mathcal{L}}}$
(overline indicates complex conjugate). Clearly the vacuum $|0\rangle\langle0|$
is a steady state (with eigenvalue zero). We use the following notation
for the Hilbert-Schmidt scalar product in $\mathcal{V}$: $\lb x|y\rb=\Tr(x^{\dagger}y)$
and use the identification $|j\rb\leftrightarrow|0\rangle\langle j|$
for $j=1,\ldots,N$ which defines a basis of $\mathcal{V}_{0,1}$.

According to Eq.~\eqref{eqn: coherence} we need the matrix element
$[\rho(t)]_{0,1}=\langle0|\rho(t)|1\rangle=\lb1|\rho(t)\rb$. Because
of the block-structure of the Lindbladian one obtains $[\rho(t)]_{0,1}=\lb1|e^{t\mathcal{L}}|\rho(0)\rb=\lb1|e^{t\tilde{\mathcal{L}}}|\tilde{\rho}(0)\rb,$where
we indicated with $\tilde{\rho}(0)$ the projection of $\rho(0)$
to $\mathcal{V}_{0,1}$ according to the above direct sum decomposition
of $\mathcal{V}$. Note that if the qubit is initialized in the state
$\alpha_{0}|\uparrow\rangle+\beta_{0}|\downarrow\rangle$, we have
$\tilde{\rho}(0)=\overline{\alpha_{0}}\beta_{0}|0\rangle\langle1|$
or equivalently $|\tilde{\rho}(0)\rb=\overline{\alpha_{0}}\beta_{0}|1\rb$.
In the following we will always consider $\overline{\alpha_{0}}\beta_{0}=1/2$,
i.e.~maximal initial coherence, such that 
\begin{equation}
\mathcal{C}(t)=\left|\lb1|e^{t\tilde{\mathcal{L}}}|1\rb\right|.\label{eq:coherence_2}
\end{equation}
As usual we can identify $\mathcal{V}_{0,1}\simeq\mathbb{C}^{N}$,
and the Hilbert-Schmidt scalar product carries over to the $\ell^{2}$
scalar product. We also use the the norm $\left\Vert x\right\Vert =\sqrt{\lb x|x\rb}$
for $x\in\mathcal{V}_{0,1}$ and the induced norm for elements of
$L(\mathcal{V}_{0,1})$. Since the basis $|j\rb$ is orthonormal,
Hilbert-Schimdt adjoint simply corresponds to transposition and complex
conjugation in this basis. With these identifications the setting
resembles that of standard one-particle quantum mechanics, with the
important difference that operators are not Hermitian. For example,
for the case of a single qubit, $M=1$, interacting with a single
cavity and cavities connected on a linear geometry $J_{i}=J_{i,i+1}$
(see Figure \ref{fig:impurity} for a schematic picture), the matrix
$\tilde{\mathcal{L}}$ becomes

\begin{equation}
\tilde{\mathcal{L}}=-i\left(\begin{array}{ccccc}
\omega_{1}^{0} & \kappa & 0 & \cdots & 0\\
\kappa & \omega_{1}-i\frac{\Gamma_{1}}{2} & J_{1} & \cdots & 0\\
0 & J_{1} & \omega_{2}-i\frac{\Gamma_{2}}{2} & \cdots & 0\\
\vdots & \vdots & \vdots & \ddots & J_{K}\\
0 & 0 & 0 & J_{K} & \omega_{K}-i\frac{\Gamma_{K}}{2}
\end{array}\right)\equiv-i\mathsf{H},\label{eqn: H}
\end{equation}
where we also defined the matrix $\mathsf{H}$ which is a non-Hermitian
generalization of a tight-binding chain.

\textbf{Remark. }The $\ell^{2}$ scalar product (and corresponding
norm) in $\mathcal{V}_{0,1}$ is natural in that, via Hilbert-Schmidt,
allows to move from Schrödinger to Heisenberg representation. However
in this setting, the $\ell^{2}$ moduli square \emph{are not probabilities}.
Conservation of quantum-mechanical probabilities is enforced by the
complete positivity and trace preserving property of the full map
$e^{t\mathcal{L}}$ for $t\ge0$. Trace conservation in turn implies
$\lb\1|\mathcal{L}=0$, where $\lb\1|$ corresponds to the identity
operator on the Hilbert space $\mathcal{V}$. This property, however,
does not carry over to the restricted generator $\tilde{\mathcal{L}}$.
What can still be said is that the eigenvalues of $\tilde{\mathcal{L}}$,
since they are a subset of those of $\mathcal{L}$, fulfill $\mathrm{Re}(\lambda)\le0$.

In general $\mathcal{C}(t)$ will decay in time starting form its
maximum value 1 at $t=0$. From Eq.~\eqref{eq:coherence_2} we realize
that our goal is to make a particular matrix element of the restricted
evolution $e^{t\tilde{\mathcal{L}}}$, have large absolute value for
possibly large times. In fact, ideally we would like: i) $\tilde{\mathcal{L}}|1\rb=\lambda_{1}|1\rb$
and ii) $\mathrm{Re}(-\lambda_{1})=0$. Both of these conditions can
be trivially achieved simply setting $\kappa_{l,1}=0$, $\forall l$.
However this entirely decouples the qubit from the rest of the network
which means one does not have a way to address the qubit anymore -
in fact experimenters generally try to \emph{increase} the qubit-mode
coupling. In view of this we replace the two conditions above with
the more physical requirements, i') $\tilde{\mathcal{L}|}1\rb\approx\lambda_{1}|1\rb$
and ii') $\mathrm{Re}(-\lambda_{1})$ as small as possible.

Condition ii') (that there exist an eigenvalue of $\tilde{\mathcal{L}}$
with almost zero real part) resembles the condition for having an
approximate zero mode familiar in (Hermitian) topological insulators.
More generally, in a linear geometry, a way to fulfill conditions
i') and ii') is to find, approximate, non-Hermitian, topological zero
mode of $\tilde{\mathcal{L}}$. Non-Hermitian generalization of topological
insulators have been studied to some extent (see e.g., \cite{rudner_topological_2009,esaki_edge_2011,zeuner_observation_2015,leykam_edge_2017}).
In particular we will be concerned with finite size systems which
have not been discussed in the literature so-far. Before turning to
topological models let us first consider what seems to be the simplest
geometry.

\section{Single impurity}

\begin{figure}
\centering{}\includegraphics[width=0.45\textwidth]{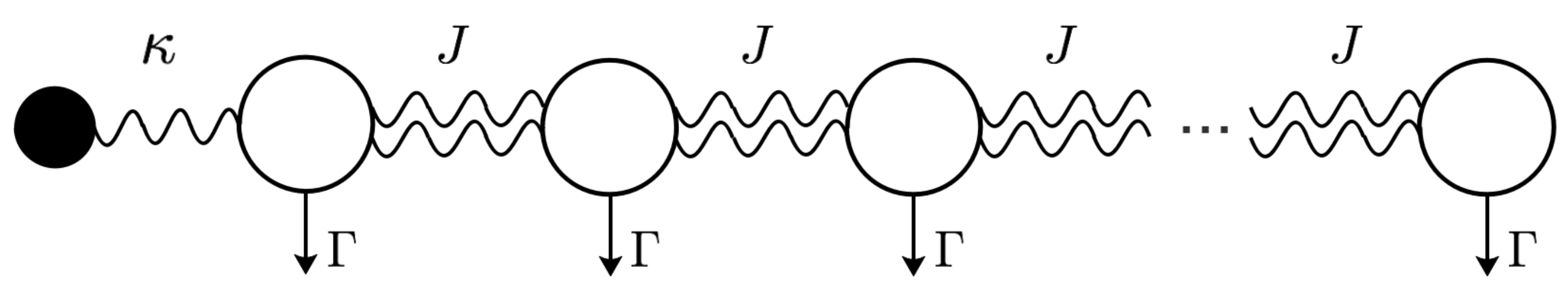} \caption{The ``single impurity model'': a qubit in a cavity connected to
a linear array of cavities. \label{fig:impurity} }
\end{figure}

The simplest case is that of linear geometry with a single impurity
(see Fig.~\eqref{fig:impurity}), i.e.~we set $J_{i}=J$, $\Gamma_{i}=\Gamma$
and also $\omega_{i}=\omega_{1}^{0}$ for all $i$ (no detuning) in
Eq.~(\ref{eqn: H}): 
\begin{equation}
\mathsf{H}=\left(\begin{array}{ccccc}
0 & \kappa & 0 & \cdots & 0\\
\kappa & \frac{i\Gamma}{2} & J & \cdots & 0\\
0 & J & \frac{i\Gamma}{2} & \cdots & 0\\
\vdots & \vdots & \vdots & \ddots & J\\
0 & 0 & 0 & J & \frac{i\Gamma}{2}
\end{array}\right),\label{eq:H_imp}
\end{equation}
where $\mathsf{H}$ has been transformed to the rotating frame of
frequency $\omega_{1}^{0}$. This is a non-Hermitian generalization
of a single impurity in a tight binding chain \cite{economou_greens_2006}.
For $N=3$ this model has been investigated in \cite{sete_quantum_2015,man_cavity-based_2015},
where it was established that adding one auxiliary cavity to a dissipative
optical cavity coupled to a qubit can significantly increase the coherence
time of the qubit. An equation for the eigenvalues can be found using
the techniques to diagonalize tridiagonal matrices. The eigenvalues
of the matrix \eqref{eq:H_imp} $\tilde{\mathcal{L}}$ can be written
as $\lambda_{k}=-i2J\cos(k)-\Gamma/2$, where $k$ is a (possibly
complex) quasi-momentum that satisfies the following equation 
\begin{equation}
\left[2\cos(k)+ia\right]\sin(kN)-\beta^{2}\sin(k(N-1))=0,
\end{equation}
where $a=\Gamma/(2J)$, $\beta=\kappa/J$. In order to look for a
localized state we look for a solution of the above equation with
complex $k=x+iy$. Essentially the localization length is given by
$\zeta=y^{-1}.$ More details are provided in Appendix \ref{sec:Single-impurity-case}.
Neglecting terms of order $O\left(e^{-N\left|y\right|}\right)$ the
eigenvalues of $\tilde{\mathcal{L}}$ of such localized modes are
given by 
\begin{equation}
\lambda_{\pm}=-\frac{4\kappa^{2}}{\Gamma\pm\sqrt{16(J^{2}-\kappa^{2})+\Gamma^{2}}}+O\left(e^{-N\left|y\right|}\right).\label{eq:lam_pm}
\end{equation}
This formula is valid in regions where $\mathrm{Re}(\lambda_{\pm})<0$.
Because the wave vector $k$ is complex, a plane wave trial solution
will decay like $e^{-yn}=e^{-n/\zeta}$ which defines the localization
length $\zeta$. In such cases the localization length is given by
\begin{equation}
\zeta=1/\ln\left|\frac{4J}{\Gamma\pm\sqrt{16(J^{2}-\kappa^{2})+\Gamma^{2}}}\right|.
\end{equation}

\begin{figure}
\centering{}\includegraphics[width=0.48\textwidth]{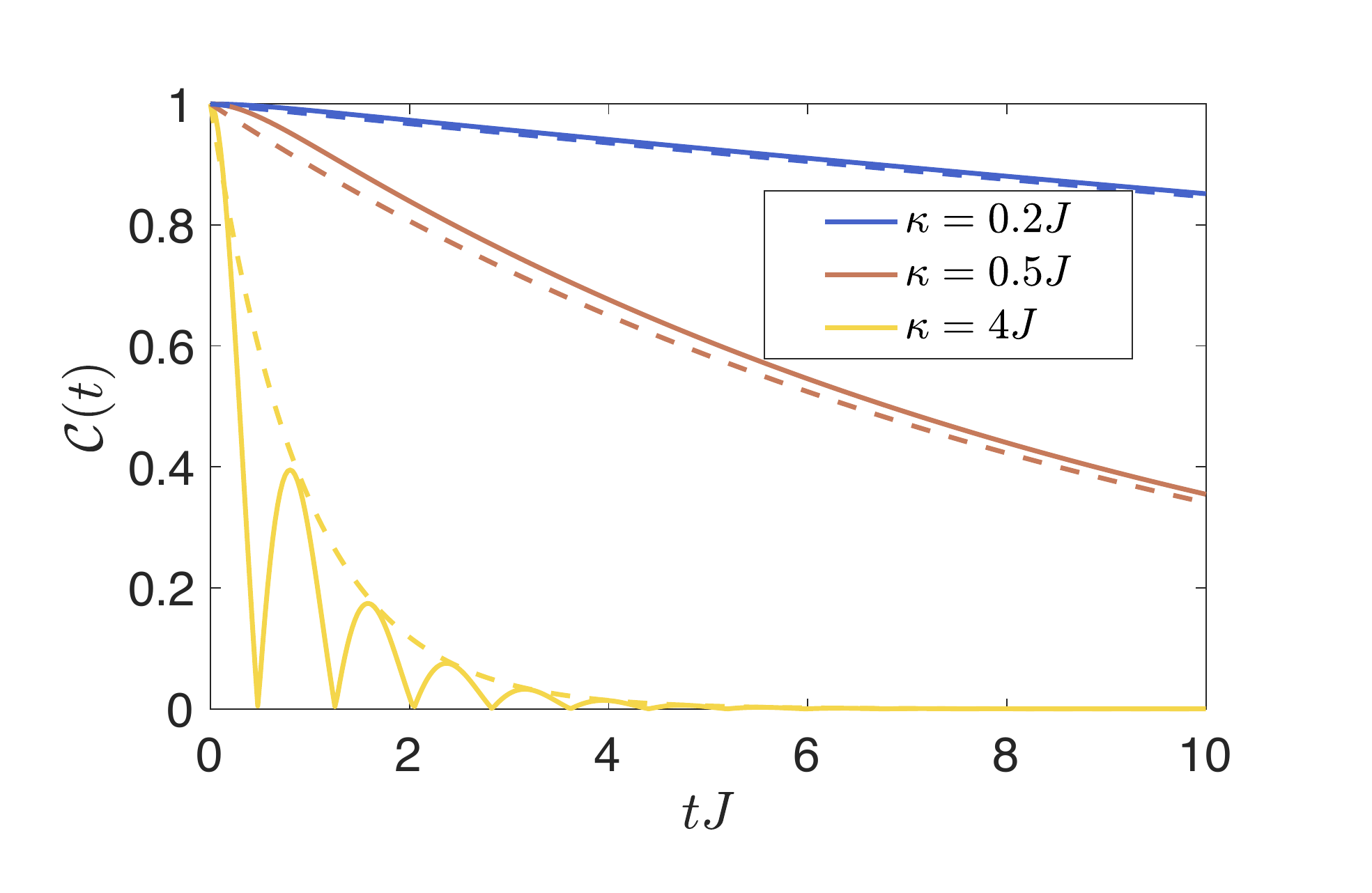}
\caption{(Color online) Behavior of the coherence for the single impurity model.
Here and in the following we compute Eq.~(\ref{eq:coherence_2})
by numerical diagonalization of the corresponding reduced Lindbladian.
Continuous lines are numerical simulation and dashed lines are analytical
approximation of Eqns.~(\ref{eq:C_imp}-\ref{eq:tau_imp}). Dissipation
is fixed to $\Gamma=4J$. The results for $N=4$ are indistinguishable
from those at $N=400$. \label{fig:COH_impurity} }
\end{figure}

For $\kappa/\Gamma$ small (strong dissipative regime), using a perturbative
argument (more details in Appendix \eqref{sec:Conditions-to-optimize}),
one can show that the coherence has approximately the form of a single
exponential decay $e^{-t/\tau_{0}}$, with $\tau_{0}^{-1}=2\kappa^{2}/\Gamma$.
Using Eq.~(\ref{eq:lam_pm}) the eigenvalue connected with $\tau_{0}^{-1}$
is $\lambda_{+}$. By continuity, we can now we can use the expression
for the localized mode outside from the strict perturbative region.
In other words we have 
\begin{align}
\mathcal{C}(t) & \approx e^{-t/\tau}\label{eq:C_imp}\\
\tau & =\mathrm{Re}\Big[\frac{\Gamma+\sqrt{16(J^{2}-\kappa^{2})+\Gamma^{2}}}{4\kappa^{2}}\Big].\label{eq:tau_imp}
\end{align}
The above equations are extremely accurate in the region of small
$\kappa$ but surprisingly are quite accurate also for large $\kappa$.
Increasing $\kappa$ one starts observing non-Markovian oscillations\footnote{Obviously a Lindblad master equation that ignores non-Markovian effects
between the system and the bath is perfectly able to encompass non-Markovian
features between a qubit and the rest of the system. This should be
no source of confusion.} in the coherence also noted in \cite{man_cavity-based_2015} at an
energy scale of the order of $J^{2}+\Gamma^{2}/16$ (when the square
root term in Eq.~(\ref{eq:tau_imp}) becomes imaginary). In this
regime Eq.~(\ref{eq:C_imp}) describes well the envelope of the coherence.
See Fig.~\ref{fig:COH_impurity} for comparisons with numerics.

\section{Topological classification of dissipative systems\label{sec:Topological-classification-of}}

We recall here for completeness the basics of the topological classification
of models of Ref.~\cite{rudner_survival_2016} (see also \cite{esaki_edge_2011}).
Since eigenvalues are now complex, there are at least two ways to
generalize this notion to the non-Hermitian world. Namely one may
extend the role played by the Hermitian gap to either the imaginary
or the real part of the eigenvalues. Two points in parameter space
are defined to be in the same phase if the corresponding (non-Hermitian)
Hamiltonians can be smoothly connected without closing the imaginary
(resp.~real) part of the eigenvalues. For the ``imaginary-gap''
classification of Ref.~\cite{rudner_survival_2016}, according to
a generalized bulk-edge correspondence, a non-trivial phase at finite
size would have edge modes with infinite or exponentially large life-time.
Clearly this is the relevant classification in our context.

We assume a periodic linear chain with $n$ sites per unit cell such
that, in the thermodynamic limit, the Hamiltonian is given by \emph{$\mathsf{H}=\oint dk/(2\pi)\sum_{\alpha,\beta}\mathsf{H}_{\alpha,\beta}(k)|k,\alpha\rb\lb k,\beta|$}
and we simply need to focus on the $n\times n$ Bloch matrix $\mathsf{H}(k)$.
The dissipation has the special form shown in Sec.~\ref{sec:Mapping-to-a}
which consists of imaginary terms on the diagonal (of negative imaginary
part). Without constraint such models are topologically trivial if
the number of leaky sites per cell is greater than one \cite{rudner_survival_2016}.
We then focus on the case where there is only one leaky site per cell.
As shown in \cite{rudner_survival_2016}, any such $\mathsf{H}(k)$
that does not admit a dark state can be written in the following way
\begin{equation}
\mathsf{H}(k)=\left(\begin{array}{cc}
U(k) & 0\\
0 & 1
\end{array}\right)\left(\begin{array}{cc}
\tilde{h}(k) & \tilde{v}_{k}\\
\tilde{v}_{k}^{\dagger} & \Delta(k)-i\Gamma
\end{array}\right)\left(\begin{array}{cc}
U(k)^{\dagger} & 0\\
0 & 1
\end{array}\right),
\end{equation}
where $\tilde{h}(k)$ is an $(n-1)\times(n-1)$ diagonal matrix with
real eigenvalues, $U(k)$ is an $(n-1)\times(n-1)$ unitary matrix
that diagonalizes $\tilde{h}(k)$ and also makes the $(n-1)$ dimensional
vector $\tilde{v}_{k}$ real and positive. Any $U(k)$ satisfying
the above criteria can be chosen without affecting the following result.
In Ref.~\cite{rudner_survival_2016} it is further shown that the
winding number of $\mathsf{H}$ then reduces to the winding number
of $U(k)$ which is given by 
\begin{equation}
W=\oint\frac{dk}{2\pi i}\partial_{k}\ln\det(U(k)).\label{eq:W}
\end{equation}
From what we have said, in a non-trivial topological phase, at finite
size one expects to observe dark states localized at the edges. Such
a dark (or quasi-dark) state $|\xi\rb$ fulfills $\tilde{\mathcal{L}}|\xi\rb=\lambda|\xi\rb$
with $\mathrm{Re}(\lambda)\simeq0$. However, given the structure
of the space $\mathcal{V}_{0,1}$ all such states are e.g.~traceless.
Hence these are not strictly quantum states, they are in fact \emph{off-diagonal
elements} of a quantum state. In the quantum-chemistry community these
are sometimes called \emph{coherences}.

We would like to conclude this section by reminding a general result
for completely positive maps/semigroups. We assume here finite dimensionality.
Let the Jordan decomposition of $\mathcal{L}$ be $\mathcal{L}=\sum_{k}\lambda_{k}P_{k}+D$
where $D$ is the nilpotent part. Define the projector onto the dark
states sector as 
\begin{equation}
P_{\mathrm{ds}}=\sum_{k,\mathrm{Re(\lambda_{k})=0}}P_{k}.
\end{equation}
Decomposing the Liouville space as $\1=P_{\mathrm{ds}}\oplus(\1-P_{\mathrm{ds}})$
one has $e^{t\mathcal{L}}=\mathcal{W}_{t}\oplus\mathcal{R}_{t}$ where
$\mathcal{W}_{t}$ is the part of the evolution inside the dark-state
sector: $\mathcal{W}_{t}=P_{\mathrm{ds}}\mathcal{W}_{t}=\mathcal{W}_{t}P_{\mathrm{ds}}$
and the remaining term $\mathcal{R}_{t}$ can be made as small as
one wishes in norm, by taking larger $t$. It can be shown (see Theorem
6.16 of \cite{wolf_quantum_2012}) that $\mathcal{W}_{t}$ is a unitary
evolution, more precisely $\mathcal{W}_{t}[\rho_{0}]=U_{t}\tilde{\rho}_{0}U_{t}^{\dagger}$
where the state $\tilde{\rho}_{0}$ is partly determined by the initial
state $\rho_{0}$. In other words, the time evolution inside the dark
state sector is unitary.

\section{Non-Hermitian SSH model\label{sec:Non-Hermitian-SSH-model}}

To start we consider the model given by the following non-Hermitian
generalization of the SSH Hamiltonian (for simplicity we rename all
hopping constants $J_{i}$ both for qubit-mode and mode-mode hopping)
\begin{equation}
\mathsf{H}=\left(\begin{array}{ccccc}
0 & J_{1} & 0 & 0 & 0\\
J_{1} & -i\Gamma & J_{2} & 0 & 0\\
0 & J_{2} & 0 & J_{1} & 0\\
0 & 0 & J_{1} & -i\Gamma & \ddots\\
0 & 0 & 0 & \ddots & \ddots
\end{array}\right).\label{eq:H_SSH}
\end{equation}
One may obtain an intuitive understanding of the model by considering
the periodic boundary conditions version of the above. In that case
it suffices to consider the $2\times2$ Bloch Hamiltonian 
\begin{equation}
\mathsf{H}(k)=\left(\begin{array}{cc}
0 & v_{k}\\
\overline{v}_{k} & -i\Gamma
\end{array}\right),\label{eq:hk_SSH}
\end{equation}
with $v_{k}=J_{1}+J_{2}e^{ik}$. Model (\ref{eq:hk_SSH}) is, up to
a constant term, pseudo-anti-Hermitian, in that $\tilde{\mathsf{H}}(k):=\mathsf{H}(k)+i(\Gamma/2)\1$
satisfies $\sigma^{z}\left[\tilde{\mathsf{H}}(k)\right]^{\dagger}\sigma^{z}=-\tilde{\mathsf{H}}(k)$.
Moreover $\tilde{\mathsf{H}}(k)$ is a linear combination of the matrices
$\left\{ -\sigma^{x},-\sigma^{y},i\sigma^{z}\right\} $ which span
the Lie algebra of $SU(1,1)$ ($S(1,1)$ in turn is the group of $2\times2$
complex matrices $U$ satisfying $U^{\dagger}\sigma^{z}U=\sigma^{z}$
and $\det(U)=1$). Model (\ref{eq:hk_SSH}) is then also referred
to as $SU(1,1)$ model \cite{esaki_edge_2011}. The more familiar,
Hermitian, SSH model being a $SU(2)$ model.

The eigenvalues of Eq.~(\ref{eq:hk_SSH}) are simply 
\begin{align}
\lambda_{k,\pm} & =-i\frac{\text{\ensuremath{\Gamma}}}{2}\pm\sqrt{\left|v_{k}\right|^{2}-\frac{\Gamma^{2}}{4}}\\
 & =-i\frac{\text{\ensuremath{\Gamma}}}{2}\pm\sqrt{J_{1}^{2}+J_{2}^{2}+2J_{1}J_{2}\cos(k)-\frac{\Gamma^{2}}{4}},
\end{align}
with momenta given by $k=4\pi n/N$ ($N$ even). For example, if $\Gamma^{2}/4<v_{\mathrm{min}}^{2}\equiv(J_{1}^{2}+J_{2}^{2}-2\left|J_{1}J_{2}\right|)$,
the square root term above is real and all the modes decay at a rate
$\Gamma/2$. This model admits a topological phase characterized by
a winding number according to the ``imaginary gap'' classification
of \cite{rudner_survival_2016}. The winding number $W$ Eq.~(\eqref{eq:W})
turns out to be analogous to that of the Hermitian SSH model, and
it simply counts the number of times the vector $J_{1}+J_{2}e^{ik}$
winds around the origin as $k$ moves around the Brillouin zone $[0,2\pi)$.
Consequently $W=1$ for $\left|J_{2}\right|>\left|J_{1}\right|$ while
$W=0$ for $\left|J_{2}\right|<\left|J_{1}\right|$ \footnote{It turns out that this model is topological also according to the
(real-gap) classification proposed in \cite{esaki_edge_2011}. More
precisely the shifted matrix $\tilde{\mathsf{H}}(k)$ has exactly
the same (real-gap) classification \cite{esaki_edge_2011}.}.

This picture gets modified for an open chain. Most importantly, as
a consequence of the topological character of the model and the so-called
bulk-edge correspondence, there will appear edge state(s) localized
at the boundary of the chain. The calculations are different depending
on whether $N$ is even or odd. We fix the geometry by fixing the
dissipation to act only on the even sites as in Eq.~(\ref{eq:H_SSH}).

\subsection{$N$ odd}

\begin{figure}
\centering{}\includegraphics[width=0.48\textwidth]{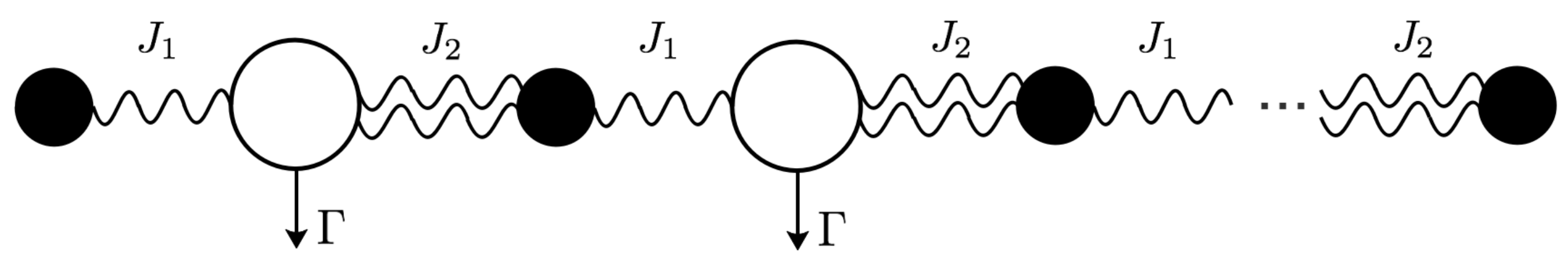} \caption{Non-Hermitian SSH model Eq.~(\ref{eq:H_SSH}) for $N$ odd. \label{fig:SSH_odd} }
\end{figure}

For $N$ odd the configuration of the bonds is given in Fig.~\ref{fig:SSH_odd}.
For $N$ odd there is always one edge state irrespective of the values
of $J_{1},J_{2}$. In this case the edge-mode has exactly zero eigenvalue
i.e., is a dark state. The edge mode is localized at the site where
the weak link is (whether it is $J_{1}$ or $J_{2}$). Clearly the
transition is at $J_{1}=J_{2}$. If $J_{1}$ is the weak link we can
write such an edge mode as 
\begin{equation}
|\xi_{L}\rb=A\left(\begin{array}{c}
e^{ik}\\
0\\
e^{3ik}\\
0\\
e^{5ik}\\
\vdots\\
e^{Nik}
\end{array}\right)\label{eq:edge_mode}
\end{equation}
where $A$ is a normalization factor. One finds that $\mathsf{H}|\xi_{L}\rb=0$
provided $J_{1}+J_{2}e^{2ik}=0$. Under this condition $|\xi_{L}\rb$
is a dark state. From this equation we see that 
\[
\left|\lb n|\xi_{L}\rb\right|^{2}=A^{2}e^{-n\delta}
\]
for $n$ odd, where $\delta\equiv\ln(|J_{2}/J_{1}|)>0$ was assumed
to be positive. Hence we call $\ell\equiv1/\ln(|J_{2}/J_{1}|)$ the
localization length of the edge mode. Fixing the normalization one
finds 
\begin{equation}
A^{2}=\frac{1-x^{2}}{x-x^{N+2}},
\end{equation}
with $x=\left|J_{1}/J_{2}\right|<1$.

The case $\left|J_{2}\right|<\left|J_{1}\right|$ can be reduced to
the previous one by a left-right symmetry transformation. Under this
transformation the dark state is mapped onto $|\xi_{R}\rb$ which
is localized at the opposite end of the chain.

Recalling the result for the periodic case one sees that, in general,
the other, non-localized, modes decay on a relaxation time-scale given
by $\tau_{\mathrm{relax}}\approx\Gamma^{-1}O(1)$. Coming to the behavior
of the coherence we see that, after a time $\tau_{\mathrm{relax}}$
all but the mode $|\xi_{L}\rb$ will have decayed. Hence the coherence,
for $t>\tau_{\mathrm{relax}}$, is approximately given by 
\begin{align}
\mathcal{C}(t) & =\left|\sum_{k}e^{\lambda_{k}t}\lb1|P_{k}1\rb\right|\nonumber \\
 & \approx\left|\lb1|\xi_{L}\rb\lb\xi_{L}|1\rb\right|=\left|\lb\xi_{L}|1\rb\right|^{2}\nonumber \\
 & =\frac{1-x^{2}}{1-x^{N+1}}.\label{eq:coh}
\end{align}
Note that, since $x<1$, this is a decreasing function of $N$. The
largest value with $N>1$, odd, is obtained for $N=3$.

For $\left|J_{2}\right|<\left|J_{1}\right|$ the role of $|\xi_{L}\rb$
and $|\xi_{R}\rb$ are reversed. Hence now the dark state is localized
at the end of the chain. After a time $\tau_{\mathrm{relax}}$ the
coherence drops to a value $\mathcal{C}(t)\simeq\left|\lb\xi_{R}|1\rb\right|^{2}=\left|\lb\xi_{L}|N\rb\right|^{2}=z^{N-1}(1-z^{2})(1-z^{N+1})^{-1}$,
where $z$ is now $z=\left|J_{2}/J_{1}\right|$, i.e.~an exponentially
small value. The two asymptotic expressions are in fact the same and
can be combined in a single expression valid for all $J_{1},J_{2}$
\begin{equation}
\mathcal{C}(t)\simeq\begin{cases}
J_{2}^{N-1}\frac{J_{2}^{2}-J_{1}^{2}}{J_{2}^{N+1}-J_{1}^{N+1}} & J_{1}\neq J_{2}\\
\frac{2}{N+1} & J_{1}=J_{2}
\end{cases}.\label{eq:coh_SSH_odd}
\end{equation}

\begin{figure}
\centering{}\includegraphics[clip,width=0.46\textwidth]{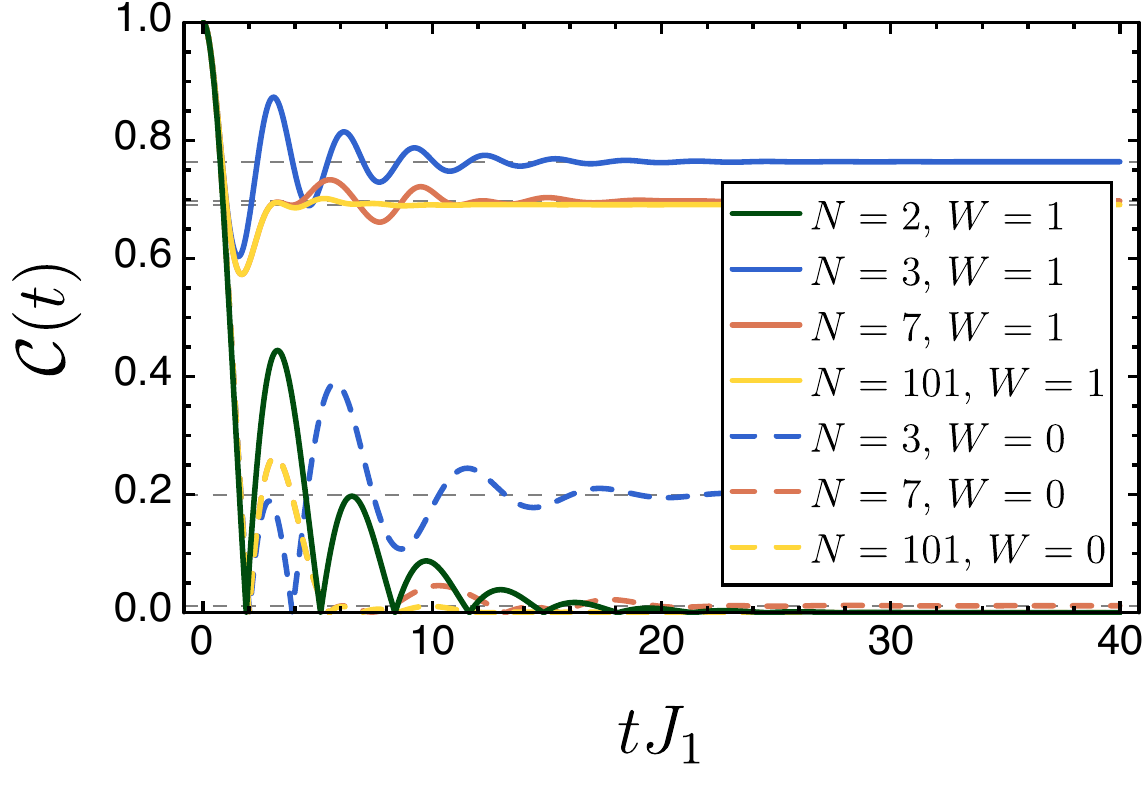}
\caption{(Color online) Behavior of the coherence in the non-Hermitian SSH
model with an odd number of sites. Continuous lines are results in
the topological phase ($W=1)$ with parameters $J_{1}=1,\,J_{2}=1.8$
and $\Gamma=0.5$. Dashed lines are for the topologically trivial
phase ($W=0$ , $J_{1}=1,\,J_{2}=0.5$ $\Gamma=0.5$). The thin dashed
lines are the asymptotic values given by Eq.~(\ref{eq:coh_SSH_odd}).
The qubit has infinite lifetime for all values of parameters, but
the asymptotic coherence is exponentially small in the topologically
trivial region. The intrinsic coherence lifetime of the qubit $(N=2)$
is added for comparison. We observe that the lattice of cavities with
$W=1$ vastly improves the lifetime of the coherence. \label{fig:COH_SSH_odd} }
\end{figure}

To summarize, for $N$ odd there is always an exact localized dark
state for all values of parameters and consequently an infinite lifetime
of the coherence's qubit. However, in the topologically trivial phase
$W=0$ ($\left|J_{1}\right|>\left|J_{2}\right|$) the edge mode is
localized at the opposite end of the chain, and the asymptotic value
of the coherence is exponentially small. The numerical simulations
confirm that a non-trivial topological winding number has a strong
effect on the coherence time of the qubit, as illustrated on Fig.~\ref{fig:COH_SSH_odd}.

To connect with the previous discussion we see that, in general we
satisfy the requirement ii') (there is an eigenmode with $\mathrm{Re}(\lambda)=0$),
but not necessarily i') . In other words, in general $|\xi_{L}\rb\lb\xi_{L}|$
is not close to $|1\rb\lb1|$. We progressively enter this regime
when the localization length becomes very short (or $\delta$ very
large). Clearly this happens when $\left|J_{2}\right|\gg\left|J_{1}\right|$.

\subsection{$N$ even}

\begin{figure}
\centering{} \includegraphics[width=0.48\textwidth]{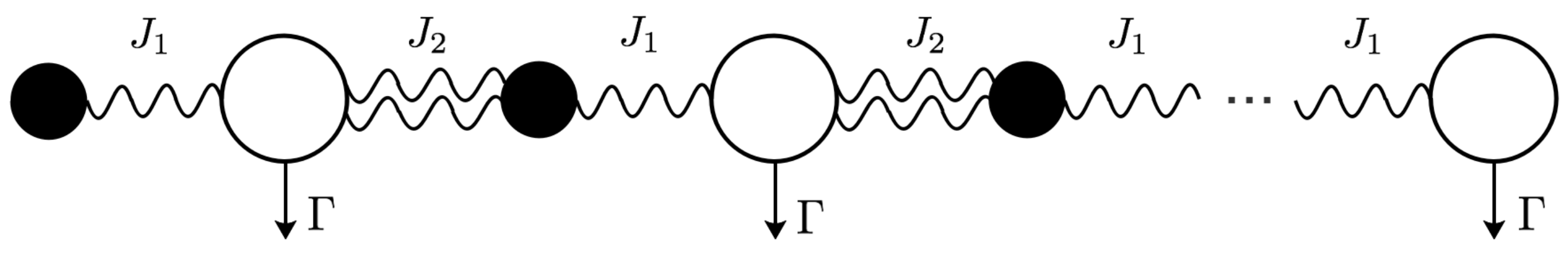} \caption{Non-Hermitian SSH model Eq.~(\ref{eq:H_SSH}) for $N$ even. \label{fig:SSH_even} }
\end{figure}

For $N$ is even the configuration of the links is depicted in Fig.~\ref{fig:SSH_even}.
When $N$ is even, $|\xi_{L}\rb$ of Eq.~(\eqref{eq:edge_mode})
does not satisfy the last row of the eigenvalue equation but rather
one has $\mathsf{H}|\xi_{L}\rb=J_{1}e^{ik(N-1)}|N\rb$. This is consistent
with our expectation of an exponentially small eigenvalue. The exact
diagonalization of the model can be found in \cite{kuznetsova_exact_2006}
(see also \cite{feldman_exact_2005,campos_venuti_long-distance_2007}).
For $N$ even edge modes appear for $d\equiv J_{2}/J_{1}>1+2/N$.
This is an interesting effect as one can in principle enter the topologically
non-trivial phase for fixed values of the parameters by only changing
$N$. The eigenvalues of the edge modes are given by \cite{kuznetsova_exact_2006}
\begin{equation}
\lambda_{\pm}=-i\frac{\text{\ensuremath{\Gamma}}}{2}\pm\sqrt{J_{1}^{2}+J_{2}^{2}+2J_{1}J_{2}\cosh(y)-\frac{\Gamma^{2}}{4}}
\end{equation}
where $y$ satisfies 
\begin{equation}
\sinh(\frac{N}{2}y)=x\sinh\left[(\frac{N}{2}+1)y\right].\label{eq:eqy}
\end{equation}
For $N$ large the solution of Eq.~(\ref{eq:eqy}) approaches $e^{y}=d$.
Up to first order in $d^{-N}$ one obtains that the solution of Eq.~(\ref{eq:eqy})
is 
\begin{equation}
e^{y}=d+d^{-N}\left(d^{-1}-d\right)+O(d^{-2N}).
\end{equation}
Plugging the above into Eq.~(\ref{eq:eqy}) one finds 
\begin{align}
\lambda_{+} & =-i\frac{J_{1}^{2}}{\Gamma}d^{-N}\left(d^{-1}-d\right)^{2}\label{eq:lam_plus}\\
\lambda_{-} & =-i\Gamma+i\frac{J_{1}^{2}}{\Gamma}d^{-N}\left(d^{-1}-d\right)^{2}.
\end{align}
The $\lambda_{+}$ eigenvalue corresponds to the mode localized at
the first site of the chain. Moreover, even if there are two localized
modes, only one of them has exponentially large life-time in the system
size. So for $N$ even the the left edge mode has a coherence time
of $\tau_{\mathrm{coh}}=\Gamma J_{1}^{-2}d^{N}(d^{-1}-d)^{-2}$. The
$\lambda_{-}$ eigenvalue corresponds to edge mode localized at the
end of the chain, with fastest decay time.

In order to compute the coherence we need the first component of the
edge mode $|\xi_{+}\rb$. It turns out that (see \cite{kuznetsova_exact_2006})
\begin{equation}
\left|\lb1|\xi_{+}\rb\right|^{2}=\frac{4\sinh^{2}(Ny/2)}{\left[\frac{\sinh[(N+1)y]}{\sinh(y)}-(N+1)\right]}\frac{\lambda_{+}+i\Gamma}{2\lambda_{+}+i\Gamma}.
\end{equation}
Since $\lambda_{+}$ is exponentially small, the last fraction is
exponentially close to 1 and can be evaluated up to $d^{-N}$ using
Eq.~(\ref{eq:lam_plus}). For the remaining terms we plug in the
asymptotic value $y=\ln(d)$ and obtain 
\begin{align}
\left|\lb1|\xi_{+}\rb\right|^{2} & =\frac{\left(1+\frac{J_{1}^{2}}{\Gamma^{2}}x^{N}\left(x-x^{-1}\right)^{2}\right)}{(1-x^{2})^{-1}-x^{N}(N+1)}+O(x^{2N})\nonumber \\
 & =1-x^{2}+x^{N}\left(1-x^{2}\right)^{2}\times\nonumber \\
 & \times\left((N+1)-\frac{1-x^{2}}{x^{2}}\frac{J_{1}^{2}}{\Gamma^{2}}\right)+O(x^{2N})\label{eq:coh_SSH_even}
\end{align}

\begin{figure}
\centering{}\includegraphics[clip,width=0.46\textwidth]{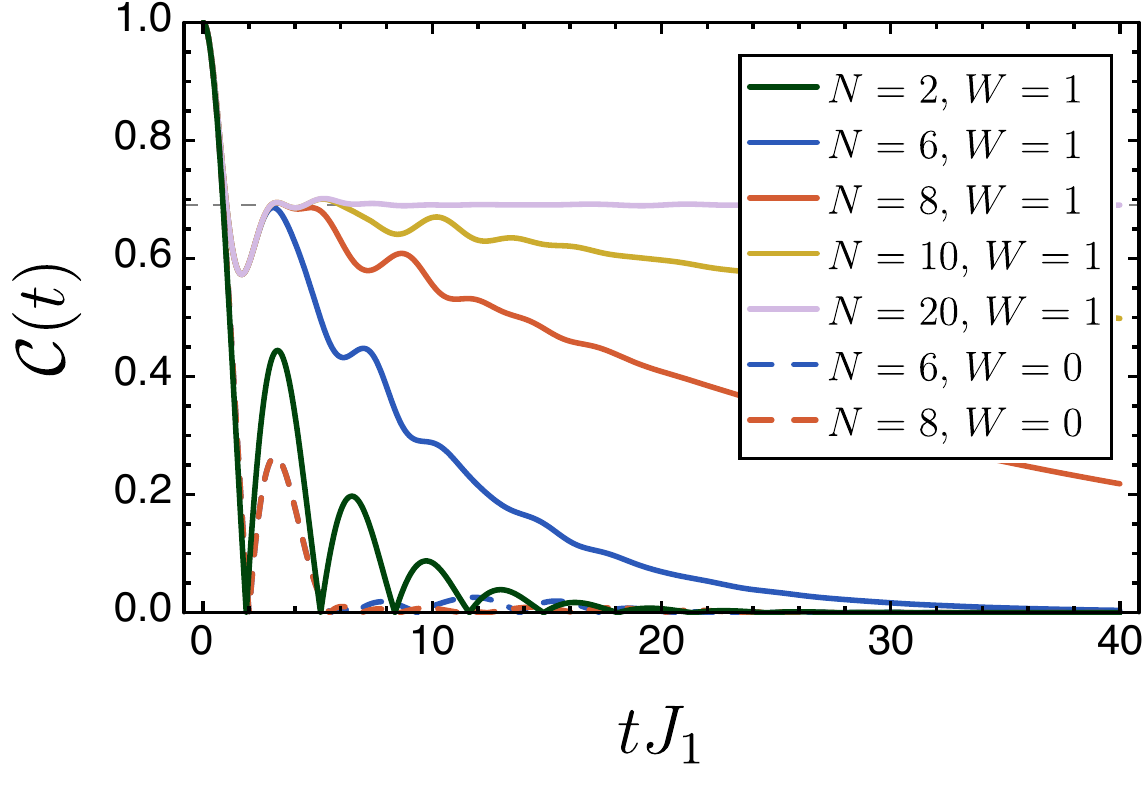}
\caption{(Color online) Behavior of the coherence in the non-Hermitian SSH
model with an even number of sites. Continuous lines are results in
the topological phase ($W=1)$ with parameters $J_{1}=1,\,J_{2}=1.8$
and $\Gamma=0.5$. Increasing $N$ has the effect of exponentially
increasing the (coherence) time-scale $\tau_{\mathrm{coh}}$ at which
the approximate dark state starts decaying. Dashed lines are for the
topologically trivial phase ($W=0$ , $J_{1}=1,\,J_{2}=0.5$ $\Gamma=0.5$).
For $N=10,\,20$ the plot is indistinguishable from that of $N=8$.
The thin dashed lines is the asymptotic value given by Eq.~(\ref{eq:coh_SSH_even}).
The intrinsic coherence lifetime of the qubit $(N=2)$ is added for
comparison. We observe that the lattice of cavities with $W=1$ vastly
improves the lifetime of the coherence. \label{fig:COH_SSH_even}}
\end{figure}

In this case the state $|\xi_{+}\rb$ is not an exact dark state and
it will start decaying at a time around $\tau_{\mathrm{coh}}$. As
for the odd case, the other states decay after a time $\tau_{\mathrm{relax}}=\Gamma^{-1}O(1)$.
Hence, whenever there is a separation of time-scales $\tau_{\mathrm{coh}}>\tau_{\mathrm{relax}}$,
one will observe a coherence of $\mathcal{C}(t)\approx\left|\lb1|\xi_{+}\rb\right|^{2}$
for times roughly in the window $t\in[\tau_{\mathrm{relax}},\tau_{\mathrm{coh}}]$.
Numerical experiments for the even case are shown in Fig.~\ref{fig:COH_SSH_even}.
In table \ref{tab:Comparison-of-exact} we show comparisons of the
numerics with the analytic expressions. For comparison, the $W=0$
case is also shown in Fig.~\ref{fig:COH_SSH_even}, where the coherence
is from bulk modes only, and the decay is given by the bulk relaxation
time $\Gamma^{-1}$.

\begin{table}
\begin{centering}
\begin{tabular}{|c|c|c|c|c|}
\hline 
$N$  & \multicolumn{2}{c|}{$\tau_{\mathrm{coh}}$} & \multicolumn{2}{c|}{$\left|\lb\xi_{+}|1\rb\right|^{2}$}\tabularnewline
\hline 
 & Exact  & Theory  & Exact  & Theory\tabularnewline
\hline 
\hline 
6  & 6.9367  & 10.9813  & 0.5355  & 0.6638\tabularnewline
\hline 
8  & 31.8117  & 35.5794  & 0.6715  & 0.6915\tabularnewline
\hline 
10  & 111.1859  & 115.2774  & 0.6888  & 0.6941\tabularnewline
\hline 
20  & $4.1153\times10^{4}$  & $4.1159\times10^{4}$  & 0.6914  & 0.6914\tabularnewline
\hline 
\end{tabular}
\par\end{centering}
\caption{Comparison of exact numerics with the approximate theoretical formulae.
Parameters are $J_{1}=1,\,J_{2}=1.8$ and $\Gamma=0.5$. \label{tab:Comparison-of-exact}}
\end{table}

\section{Three-site unit cell}

\begin{figure}
\centering{}\includegraphics[width=0.4\textwidth]{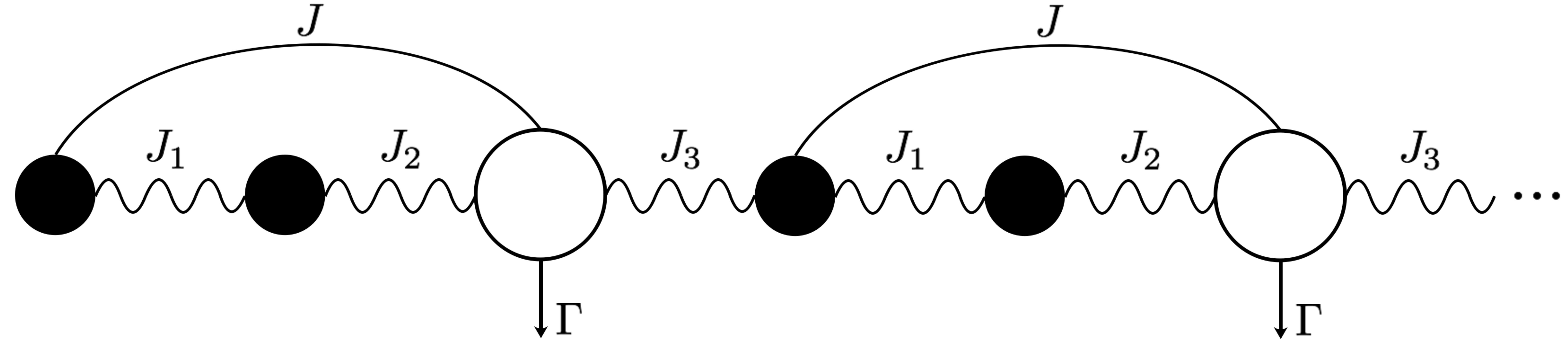}
\caption{Model (\ref{eq:H3}) with a three-site unit cell. \label{fig:three-site} }
\end{figure}

We now turn to a case where the unit cell consists of three sites.
According to the prescription of Ref.~\cite{rudner_survival_2016}
for the existence of a topological phase we consider only one leaking
site per cell. We allow for nearest neighbor hopping and also between
the first and third site in the cell (see Fig.~\eqref{fig:three-site}).
As we will see, this geometry will allow us to have topological number
of $0,1$ and $2$. The Hamiltonian is 
\begin{align}
\mathsf{H} & =\sum_{x}\Big(J_{1}|x,1\rb\lb x,2|+J_{2}|x,2\rb\lb x,3|+J_{3}|x,3\rb\lb x+1,1|\nonumber \\
 & +J|x,1\rb\lb x,3|+\mathrm{h.c.}\Big)+\nonumber \\
 & \sum_{x}\Big(\epsilon_{1}|x,1\rb\lb x,1|+\epsilon_{2}|x,2\rb\lb x,2|-i\Gamma|x,3\rb\lb x,3|\Big).\label{eq:H3}
\end{align}
For periodic boundary conditions the corresponding Bloch Hamiltonian
reads 
\[
\mathsf{H}(k)=\left(\begin{array}{ccc}
\epsilon_{1} & J_{1} & J_{3}e^{ik}+J\\
J_{1} & \epsilon_{2} & J_{2}\\
J_{3}e^{-ik}+J & J_{2} & -i\Gamma
\end{array}\right).
\]
Using Eq.~(\ref{eq:W}) it can be shown that the winding number is
given by 
\begin{multline}
W=\Theta(\left|J_{3}\right|>\left|J+J_{2}\tan(\vartheta/2)\right|)\\
+\Theta(\left|J_{3}\right|>\left|J-J_{2}\cot(\vartheta/2)\right|),\label{eq:W3_gen}
\end{multline}
where $\Theta(\mathrm{true})=1$, $\Theta(\mathrm{false})=0$ and
$\vartheta=\arccos[(\epsilon_{1}-\epsilon_{2})/\sqrt{4J_{1}^{2}+(\epsilon_{1}-\epsilon_{2})}]$.
The above quantity can assume the values $W=0,1,2$. The value $W=2$
can be obtained, for example, by taking $J_{3}$ sufficiently large.
When $W=2$ the open, finite size chain has two edge modes per end.
This gives the possibility to encode a qubit in the dark state manifold
of the model. In the following we restrict to the case $\epsilon_{2}=\epsilon_{1}=\epsilon$
for which 
\begin{equation}
W=\Theta(\left|J_{3}\right|>\left|J+J_{2}\right|)+\Theta(\left|J_{3}\right|>\left|J-J_{2}\right|).\label{eq:W3}
\end{equation}
As we can see from the above the presence of the two-sites hopping
$J$ is not necessary for having $W=2$ but it allows to have $W=1$.

\begin{figure}[h]
\centering{}\includegraphics[clip,width=0.48\textwidth]{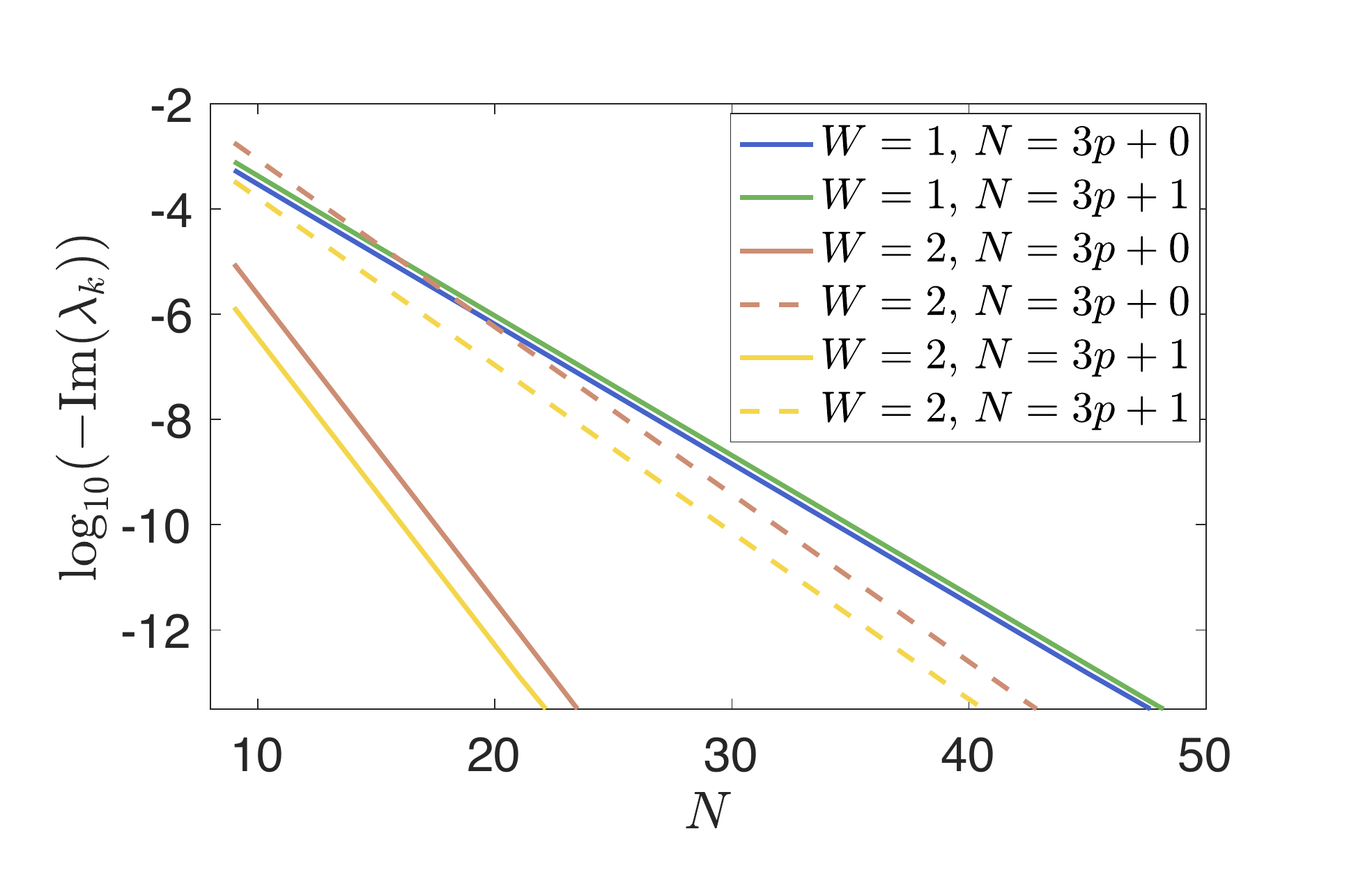}
\caption{(Color online) Scaling of the imaginary part of the eigenvalues of
the edge modes, for different topological sectors and different values
of $N\mod3$. For $N=3p+2$ we have observed always two exact dark
states($\mathrm{Im}(\lambda_{k})=0$) for all parameters values. This
simulations suggest that, for $N\mod3\protect\neq2$ there are $W$
edge modes with exponentially large life-time. Parameters are $\epsilon_{1}=\epsilon_{2}=0$,
$J_{1}=1.4,\,J_{2}=0.3,\,J=0.7$, $\Gamma=1.5$ and $J_{3}=1$ for
$W=1$ while $J_{3}=3$ for $W=2$. \label{fig:scaling}}
\end{figure}

As we have seen in section \ref{sec:Non-Hermitian-SSH-model}, at
finite size the exact number of edge modes can be a complicated function
of $N$ and the other parameters of the models. For the model of Eq.~(\ref{eq:H3}),
we have verified numerically that for $N$ mod $3=2$ there are always
(irrespective of $W$) two edge modes with imaginary part of the eigenvalues
exactly equal to zero. In other words there are always two exact dark
states. However, we have also checked that essentially only $W$ of
them are localized on the qubit site. For $N$ mod $3\neq2$ our simulations
suggest that there are $W$ edge modes with life-time exponentially
large in the system size (see Fig.~\ref{fig:scaling}). Moreover,
precisely $W$ of them are localized at the qubit site. This picture
is consistent with what we have found analytically in sec.~\ref{sec:Non-Hermitian-SSH-model}.
In other words, there are always (for all $N$) $W$ dark or quasi-dark
modes localized at the qubit site. Since, as we have seen, the behavior
of the coherence is not only dictated by the number of localized modes,
but rather by the modes\emph{ localized at the qubit}, the value of
$W$ has a strong impact on the coherence.

\begin{figure}[h]
\centering{}\includegraphics[clip,width=0.48\textwidth]{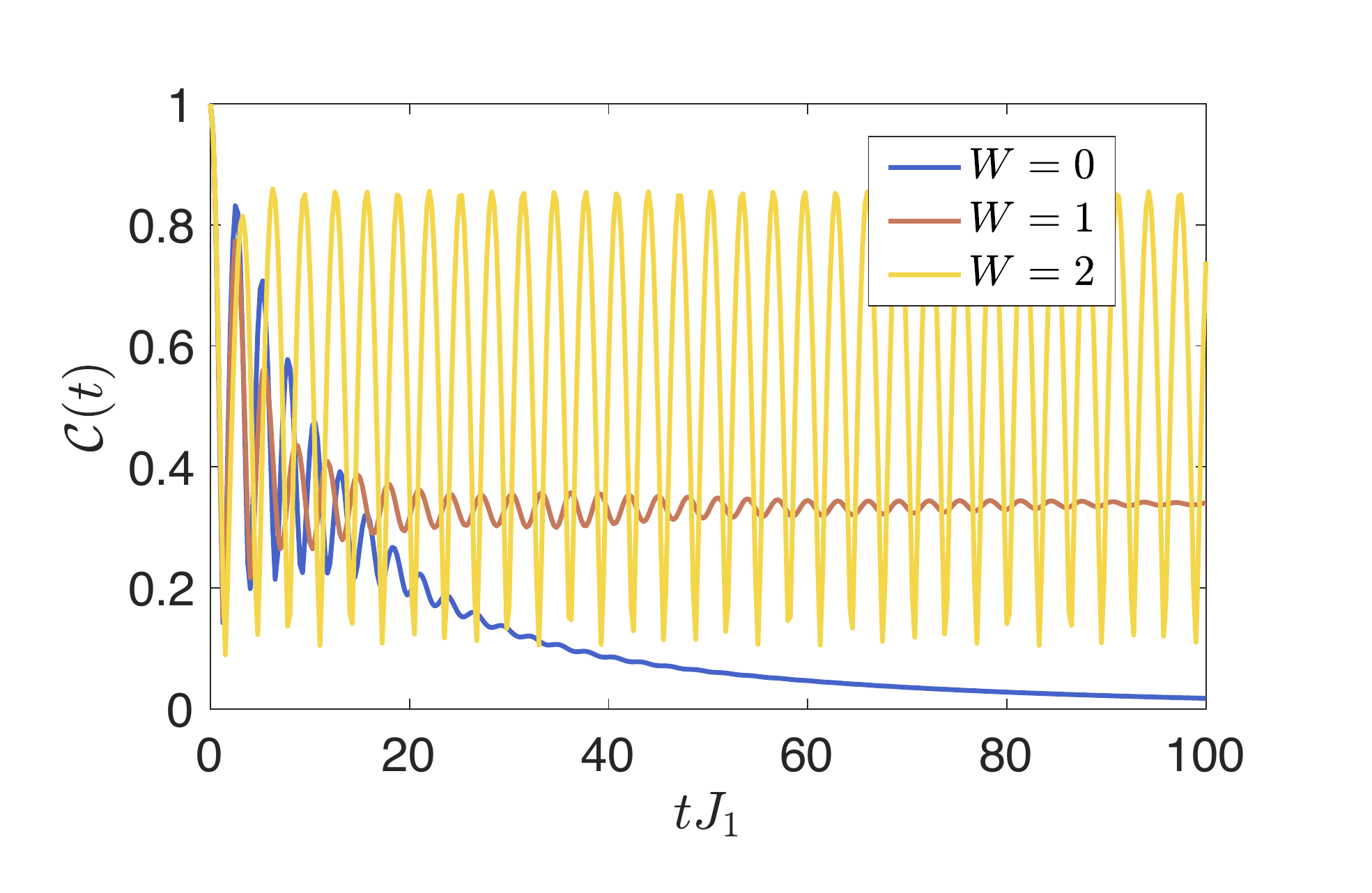}
\caption{(Color online) Behavior of the coherence in the linear chain with
three sites per cell Eq.~(\ref{eq:H3}). We chose $N=8$, but the
numerical results are not sensitive to $N$ mod 3. The winding number
can assume values $W=0,1,2$. $W$ also counts the number of edge
modes localized near the first qubit. For $W=2$ the dark state manifold
is a physical qubit and one sees Rabi oscillations in the coherence.
Parameters are $J_{1}=1,\epsilon=0,\,\Gamma=0.5,\,J_{2}=0.3,\,J=0.7$
and $J_{3}$ fixes the value of $W$: $J_{3}=0.2$ ($W=0$), $J_{3}=0.7$,
($W=1$), and $J_{3}=2$, ($W=2$). \label{fig:COH_cell3}}
\end{figure}

From what we have said so far, the behavior of the coherence of the
first qubit is now clear. For $W=0$ the coherence decays to zero
after a time $\tau_{\mathrm{relax}}=\Gamma^{-1}O(1)$ or it saturates
to an exponentially small value in $N$ if $N=3p+2$. For times $\tau_{\mathrm{relax}}\lesssim t\lesssim\tau_{\mathrm{coh}}$,
for $W=1$ it saturates to an amount given by $\mathcal{C}(t)\simeq\left|\lb1|\xi_{1}\rb\right|^{2}$
where $|\xi_{1}\rb$ is the dark state localized at the left of the
chain. For $W=2$ the coherence will oscillate between two values
in a similar way as in Rabi oscillations, $\mathcal{C}(t)\simeq\left|e^{-i\omega_{1}t}\left|\lb1|\xi_{1}\rb\right|^{2}+e^{-i\omega_{2}t}\left|\lb1|\xi_{2}\rb\right|^{2}\right|$
where $|\xi_{1,2}\rb$ are the two dark states localized at the left
with (real) eigenvalues $\omega_{1,2}$. The time-scale $\tau_{\mathrm{coh}}$
is infinite for $N=3p+2$ and exponentially large in $N$ otherwise.
A plot of the behavior of the coherence in different topological sectors
is shown in Fig.~\ref{fig:COH_cell3}.

Finally, let us comment on the long-time behavior of the full Lindbladian
evolution. For $N=3p+2$ there is an exact, non-trivial dark space
and so, for what we have said at the end of section \ref{sec:Topological-classification-of},
the evolution inside this dark space is unitary. When $N\mod3\neq2$
and $W=2$ there are two modes with life-time $\tau_{\mathrm{coh}}$
exponentially large in $N$. In this case an exact dark space sector
cannot be defined, however we have verified that the dynamics are
approximately unitary for times $t$ in the window $\tau_{\mathrm{relax}}\lesssim t\lesssim\tau_{\mathrm{coh}}$.
In this sense the term Rabi oscillations is accurate.


\section{Effect of Noise on Coherence Decay}

\begin{figure}
\centering{}\includegraphics[clip,width=0.44\textwidth]{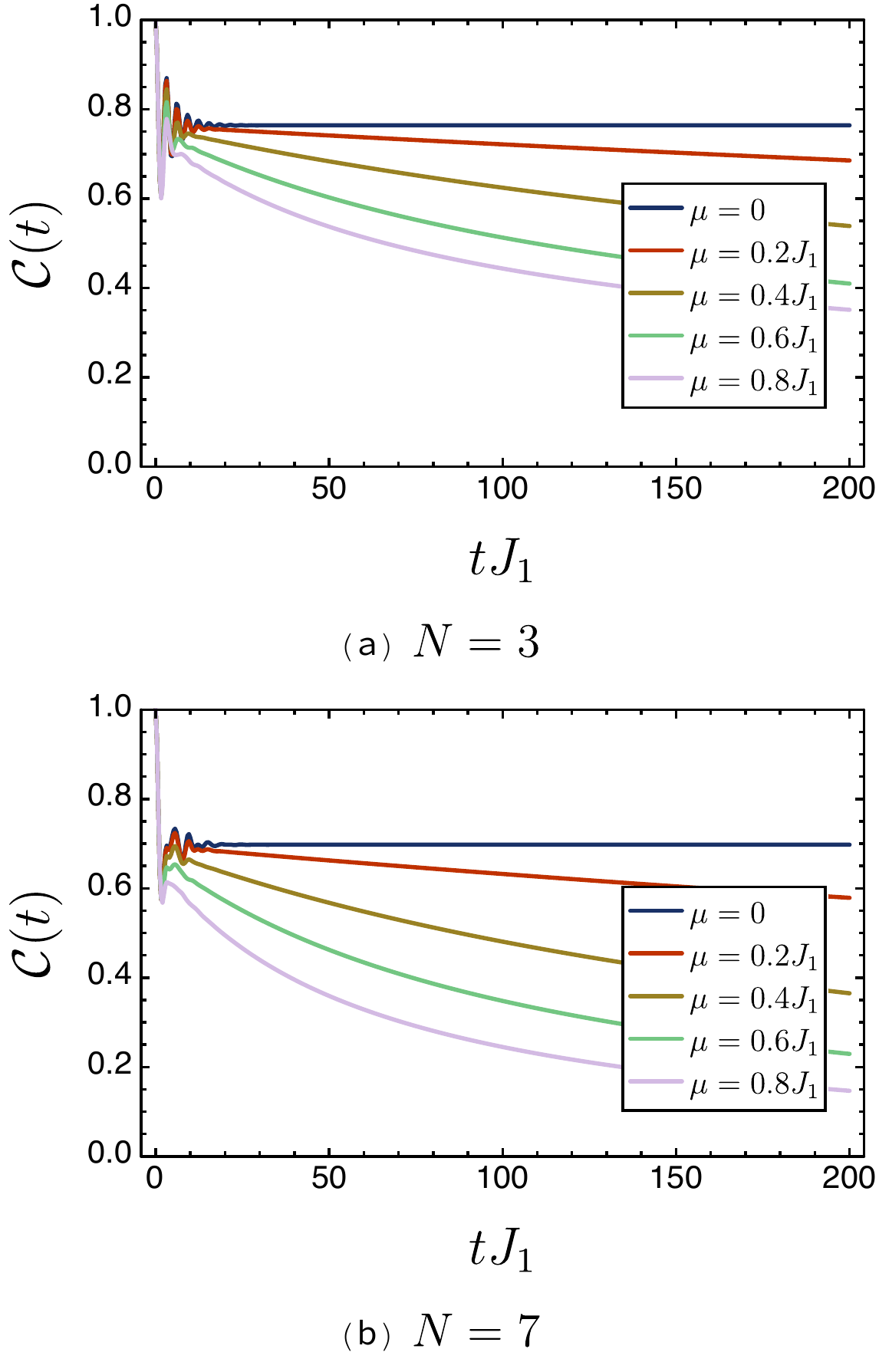}
\caption{(Color online) The effect of noise on the coherence in the non-Hermitian
SSH model with (a) $N=3$ and (b) $N=7$ respectively. We take the
$W=1$ phase, where $J_{1}=1$, $J_{2}=1.8$, $\Gamma=0.5$, and the
noise rate $\mu$ is taken between $0$ and $0.8J_{1}$. The final
result is averaged over 1000 runs of randomly generated systems with
the respective noise rates. \label{fig:SSH_noise_odd}}
\end{figure}

In this section we explore the effect of disorder on the coherence
time of our topologically protected systems. Specifically we consider
random (real) detuning of the qubits with respect to the cavity modes.
This amounts to add a diagonal term to our one-particle effective
``Hamiltonians'' with on-site ``chemical potentials'' $\mu_{i}$
where $\mu_{i}$ are i.i.d.~random variables with zero mean and uniform,
distribution in $[-\mu,\mu]$ . We compute the corresponding coherence
averaging over many (1000 in numerical simulations) realization.

\subsection{Noise Effect on the Non-Hermitian SSH Model}

We first consider the topological model of Eq.~(\ref{eq:H_SSH}).
For odd system size $N$, the topologically protected $W=1$ phase
has a true dark state, and the coherence of the qubit saturates to
a finite value (Fig.~\ref{fig:COH_SSH_odd}). We observe that the
system is quite robust against disorder (see Fig.~\ref{fig:SSH_noise_odd}).
Even for a noise strength $\mu$ comparable to the tunneling strength
$J_{1}$, the qubit's coherence remains significant over a long period
of time. In addition, a shorter chain of cavities better protects
the system against noise.

\begin{figure}
\centering{}\includegraphics[clip,width=0.44\textwidth]{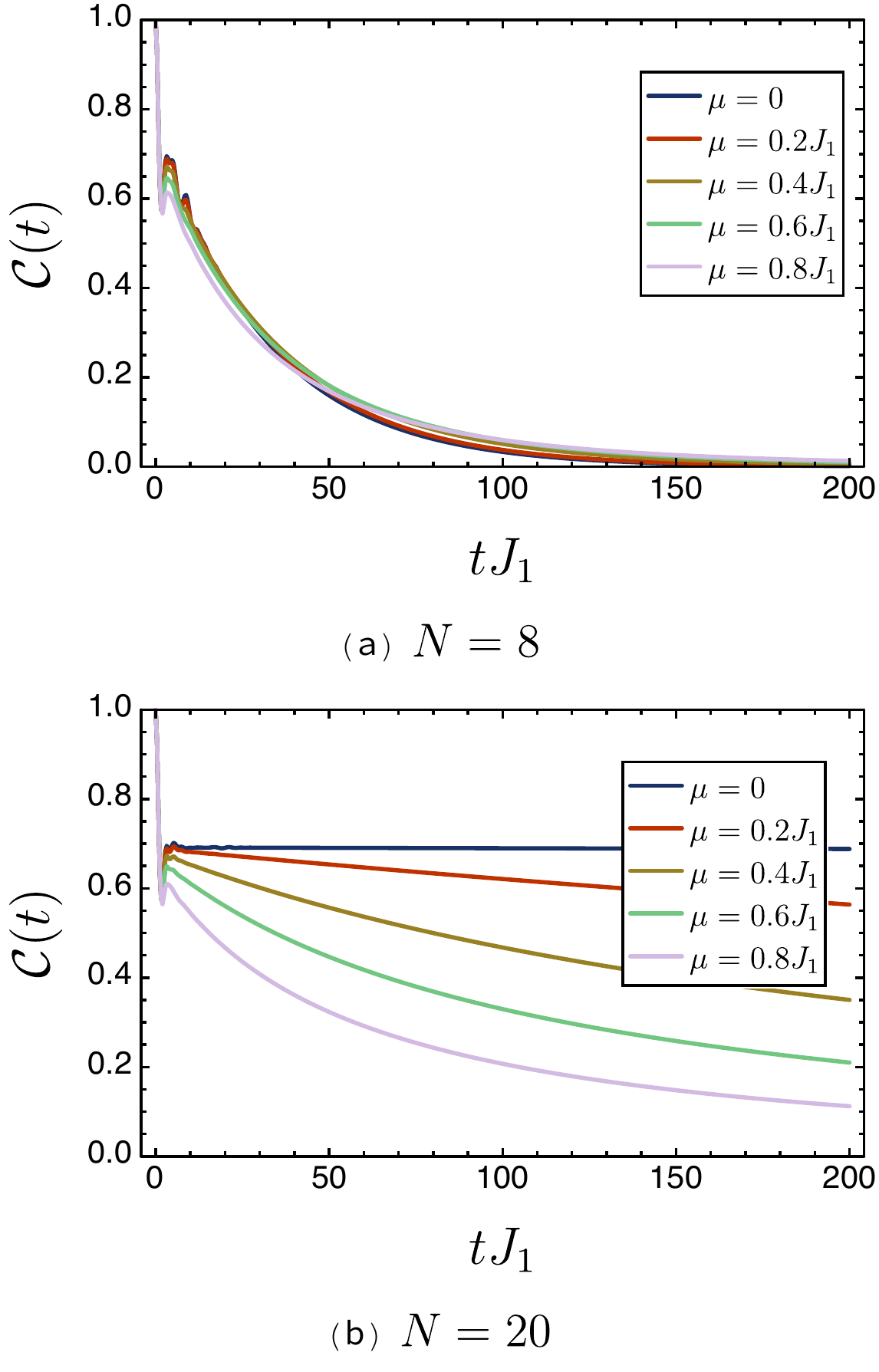}
\caption{(Color online) The effect of noise on the coherence in the non-Hermitian
SSH model with (a) $N=8$ and (b) $N=20$ respectively. We take the
$W=1$ phase, where $J_{1}=1$, $J_{2}=1.8$, $\Gamma=0.5$, and the
noise rate $\mu$ is taken between $0$ and $0.8J_{1}$. The final
result is averaged over 1000 runs of randomly generated systems with
the respective noise rates. \label{fig:SSH_noise_even}}
\end{figure}

For even system size $N$, the imaginary part of the dark state in
the $W=1$ phase is not exactly 0, but for a large enough $N$, the
qubit's coherence still saturates to a finite value for times in a
an exponentially large (in $N$) window. (Fig.~\ref{fig:COH_SSH_even})
We observe that for $N=8$ (Fig.~\ref{fig:SSH_noise_even} (a)),
where the coherence of the clean system itself decays to 0, noise
does not change the time evolution much. On the other hand, for $N=20$
(Fig.~\ref{fig:SSH_noise_even} (b)), where the coherence saturates,
the effect of noise is similar to odd $N$, and again, when the chain
of cavities gets longer, the disruptive effect of noise gets more
pronounced.

\subsection{Noise Effect on the Three-Site Unit Cell System}

Here we consider the topological model of Eq.~(\ref{eq:H3}) where
we set $\epsilon_{1}=\epsilon_{2}=0$. For this model (Fig.~\ref{fig:three-site})
there are three distinct topological phases, $W=0$, $W=1$, and $W=2$,
and the latter two protect the qubit's coherence from decaying.

\begin{figure}
\centering{}\includegraphics[clip,width=0.44\textwidth]{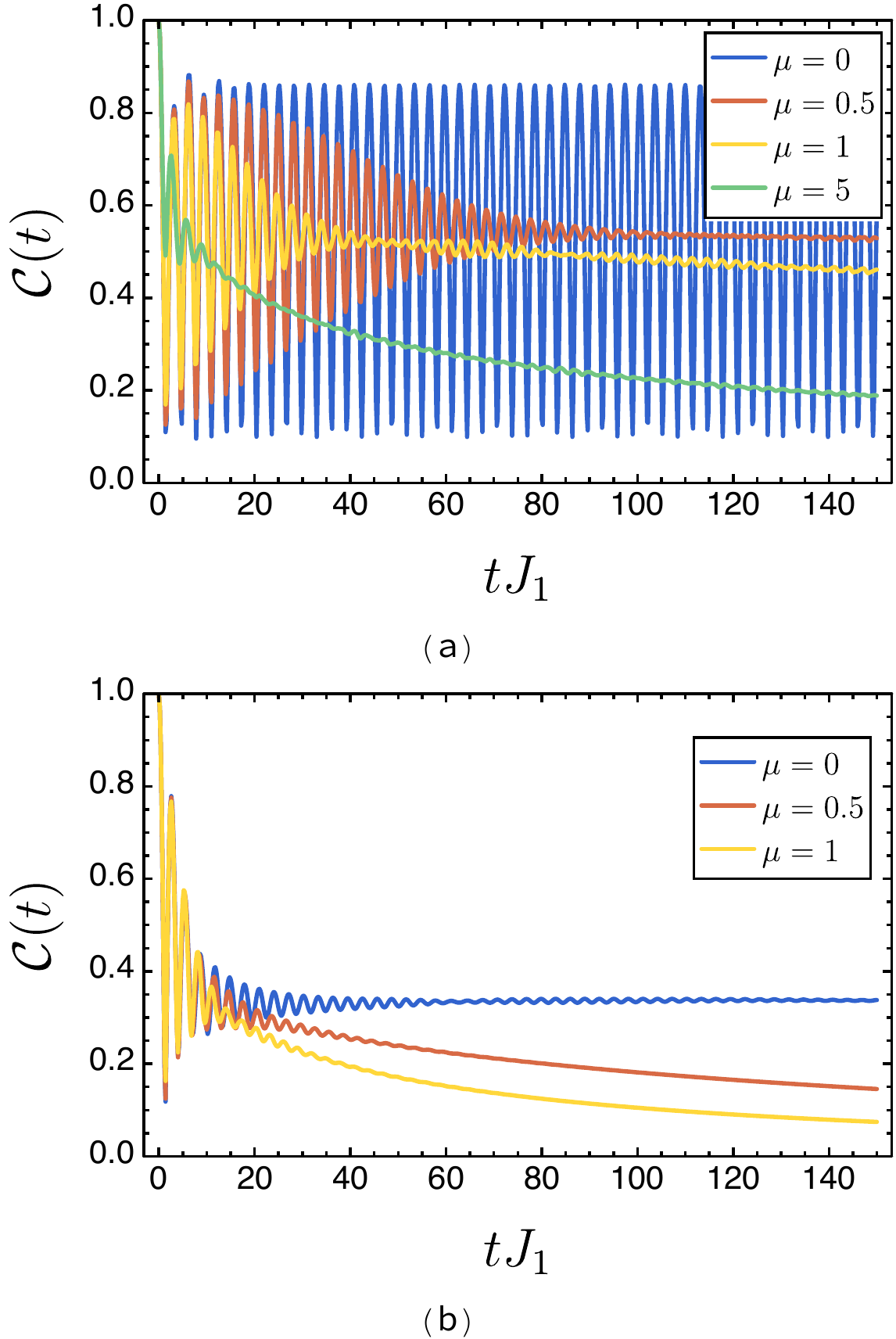}
\caption{(Color online) The effect of noise on the coherence in the three-site
unit cell model with $N=8$. (a) Phase $W=2$ with $J_{1}=1$, $J_{2}=0.3$,
$J_{3}=2$, $J=0.7$, and $\Gamma=0.5$. (b) Phase $W=1$ with $J_{1}=1$,
$J_{2}=0.3$, $J_{3}=0.7$, $J=0.7$, and $\Gamma=0.5$. The noise
rate $\mu$ is taken to be $0$, $0.5J_{1}$ and $J_{1}$. The final
result is averaged over 1000 runs of randomly generated systems with
the respective noise rates. \label{fig:three_site_noise}}
\end{figure}

For the $W=1$ phase (Fig.~\ref{fig:three_site_noise} (b)), the
effect of noise is similar to that of the $W=1$ phase of the non-Hermitian
SSH model. The coherence of the qubit no longer saturates to a finite
value, but decays to 0. We again note that the topologically protected
system is quite robust against the introduction of noise. A noise
strength comparable to the first tunneling rate ($\mu=J_{1}$) does
not decrease the coherence much even over a long time.

For the $W=2$ phase (Fig.~\ref{fig:three_site_noise} (a)), adding
detuning noise vastly alters the oscillatory behavior of the clean
system. In this case the time evolution of the coherence under noise
resembles that of the $W=1$ case, in other words the coherence saturates
to a finite value. Yet larger noise strength seems to be able to eventually
drive the system to a $W=0$-like phase where the coherence decays
to zero (see Fig.~\ref{fig:three_site_noise} (a), $\mu=5J_{1}$).
Further investigations are needed to clarify the nature of this noise-induced,
dissipative, topological phase transition \cite{campos_venuti_notitle_nodate}.

\section{Conclusions}

Non-Hermitian topological phases in a finite system permit the construction
of states whose decay time is either infinite or exponentially large
in the system size. This feature is extremely appealing from the point
of view of creating long-lived quantum bits. In this work we have
shown that networks of qubits interacting with lossy cavities may
be configured to possess non-trivial topological structure. For networks
with a simple linear geometry, we have found that localization and
long-livedness of the topological edge modes both concur to increase
dramatically the coherence of a qubit sitting at the end of the chain.
Specifically, a non-zero topological winding number $W$ results in
an exponentially long lived qubit. Although at finite size the exact
number of edge modes is a complicated function of $W$ and $N$, there
are always $W$ edge modes localized at one end of the chain. For
$W=2$ we find that the long-time dissipative, Lindbladian evolution
becomes approximately unitary, and the coherence of the qubit displays
long-lived Rabi-oscillations. In general, such long-lived, topological
edge modes, are not legitimate quantum states, but rather they are
off-diagonal elements of quantum a states or, \emph{coherences}. The
possibility of using such long-lived coherences for quantum computation
is an interesting and challenging task for future studies. 
\begin{acknowledgments}
This work was partially supported by the ARO MURI grant W911NF-11-1-0268
the DOE Grant Number ER46240 and the ERC Advanced Grant program (H.S.).
L.C.V. would like to thank Mark Rudner for useful correspondence. 
\end{acknowledgments}

 \bibliographystyle{apsrev4-1}
\bibliography{coherence_cavities}

\appendix

\section{``Single impurity'' case\label{sec:Single-impurity-case}}

Through shift and rescaling $\tilde{\mathcal{L}}=iJ\mathsf{H}'-(\Gamma/2)\1$
we are led to consider the following matrix 
\begin{equation}
\mathsf{H}'=\left(\begin{array}{ccccc}
-ia & \beta & 0 & \cdots & 0\\
\beta & 0 & 1 & \cdots & 0\\
0 & 1 & 0 & \cdots & 0\\
\vdots & \vdots & \vdots & \ddots & 1\\
0 & 0 & 0 & 1 & 0
\end{array}\right).
\end{equation}
We write the eigenvalues as $2\cos(k)$. It can be shown that $k$
satisfies the following equation (both for $N$ even and odd) 
\begin{equation}
\left[2\cos(k)+ia\right]\sin(kN)-\beta^{2}\sin(k(N-1))=0,\label{eq:eigk}
\end{equation}
and one can restrict oneself to $0<\mathrm{Re}(k)<\pi$. In order
to look for localized states we look for a complex root of Eq.~(\eqref{eq:eigk}).
Hence we set $k=x+iy$. Plugging this in the above and forgetting
terms $e^{-N|y|}$ we obtain 
\begin{equation}
\left(2\cos(k)+ia\right)=\beta^{2}\frac{\sin(k(N-2))}{\sin(k(N-1))}\simeq\beta^{2}\begin{cases}
e^{ix}e^{-y} & y>0\\
e^{-ix}e^{y} & y<0
\end{cases}\label{eq:eigs_approx}
\end{equation}

We also set $ik=q$.

Case $y>0$. The equation is 
\[
2\cosh(q)+ia=\beta^{2}e^{q}
\]
setting $z=e^{q}$ one finds 
\[
q=\ln\left((-i)\frac{a\pm\sqrt{a^{2}+4(1-\beta^{2})}}{2(1-\beta^{2})}\right)
\]
and the corresponding eigenvalues 
\begin{equation}
\lambda=-i\frac{\beta^{2}}{2\left(1-\beta^{2}\right)}\left(a\pm\sqrt{a^{2}+4(1-\beta^{2})}\right)-ia.\label{eq:eig_ab}
\end{equation}
We need to make sure that $y=-\mathrm{Re}(q)>0$. From this we obtain
$\mathrm{Re}(\ln(z))=\ln\left|z\right|<0$ or $\left|z\right|<1$.

Case $y<0$. Now the equation is 
\[
2\cosh(q)+ia=\beta^{2}e^{-q}
\]
setting $z=e^{-q}$ one finds the same equation as for $y>0$. This
means that the eigenvalues have the same from (\ref{eq:eig_ab}),
but now $y<0$ implies $\left|z\right|>1$.

Going back to the eigenvalues of $\tilde{\mathcal{L}}=iJ\mathsf{H}'-(\Gamma/2)\1$,
remembering $a=\Gamma/(2J)$ and $\beta=\kappa/J$ we get finally
\begin{equation}
\lambda_{\pm}=-\frac{4\kappa^{2}}{\Gamma\pm\sqrt{16(J^{2}-\kappa^{2})+\Gamma^{2}}}+O\left(e^{-N\left|y\right|}\right),
\end{equation}
as shown in the main text.

\section{A note on timescales\label{sec:Timescales}}

Here we would like to define a time-scale associated to the coherence
decay. This time scale should measure the time after which the coherence
has degraded to an unacceptable value. Several definition of such
(de-)coherence time are possible. For example one may take the smallest
$\tau$ such that $\mathcal{C}(\tau)=\mathcal{C}(0)-\epsilon$. According
to Eq.~(\ref{eq:coherence_2}) $\mathcal{C}(t)$ has the form $\mathcal{C}(t)=\left|\sum_{j}c_{j}e^{\lambda_{j}t}\right|$
, where $\lambda_{j}$ are (a subset of) Lindbladian eigenvalues satisfying
$\mathrm{Re}(\lambda_{j})\le0$. Let us say that one is interested
in very small $\epsilon$. In this limit the coherence time $\tau$
becomes proportional to $\epsilon$. A meaningful definition then
would be $\tau_{\mathrm{lin}}=\epsilon/\mathrm{Re}\left(-\sum_{j}c_{j}\lambda_{j}\right)$
(the name stemming from the fact that $\mathcal{C}(t)$ is approximately
linear for $t\lesssim\tau_{\mathrm{lin}}$ ). A more conservative
definition is given by the shortest time-scale associated with the
set $\left\{ \mathrm{Re}(-\lambda_{j})\right\} $ i.e.~the timescale
defined as $\tau_{\mathrm{min}}^{-1}=\max_{j}\mathrm{Re}(-\lambda_{j})$.
If $\tau_{\mathrm{min}}$ is large one is \emph{guaranteed} that the
coherence will be close to maximal for all $0\le t\lesssim\tau_{\mathrm{min}}$
for \emph{any initial state}. This is a very pleasant feature which
makes $\tau_{\mathrm{min}}$ quite attractive. Let us also define
the the \emph{slowest} decay time of $\mathcal{C}(t)$ by $\tau_{\mathrm{max}}^{-1}=\min_{j}\mathrm{Re}(-\lambda_{j})$.
Clearly $\tau_{\mathrm{max}}$ can be much larger than any meaningful
definition of coherence time \footnote{The meaning of $\tau_{\mathrm{max}}$ is that for times $t\gtrsim\tau_{\mathrm{max}}$
the coherence is guaranteed to be essentially zero for any initial
state. But we may have lost interest in $\mathcal{C}$ long before. }. Obviously all these time-scales agree if the coherence decays as
a single exponential. Quite surprisingly in all the situation we considered
in the text, we verified that indeed $\mathcal{C}(t)$ can be well
approximated with a single exponential over a wide range of $J/\Gamma$
($\Gamma$ dissipation scale and $J$ coherent energy scale, see main
text). Some arguments why this is so will be given in the next section.
In all the cases considered the coherence timescale $\tau_{\mathrm{coh}}$
defined in the main text coincides with what is commonly called Purcell
rate in the cavity QED community. Through topological protection we
are able to exponentially increase the Purcell rate.

\section{Conditions to optimize the coherence decay\label{sec:Conditions-to-optimize}}

In the following we will identify \emph{sufficient} conditions for
the requirements i') and ii'). For simplicity we assume that $\tilde{\mathcal{L}}$
can be diagonalized \footnote{This is not a serious limitation as the set of non-diagonalizable
$\tilde{\mathcal{L}}$ has measure zero in the space of parameters,
$J_{i},\Gamma_{i},\ldots$. } with spectral resolution $\tilde{\mathcal{L}}=\sum_{j}\lambda_{j}P_{j}$.
We start analyzing the following consequence of i'):

\textbf{Fact 1. }Assume i'), i.e.~$\tilde{\mathcal{L}}|1\rb=\lambda_{1}|1\rb+\epsilon|e\rb$
with $\left\Vert e\right\Vert =O(1)$. Then, up to an error $\epsilon$,
the evolution of the coherence is governed by a single exponential,
in particular 
\begin{equation}
\mathcal{C}(t)=\left|e^{t\lambda_{1}}\right|+O(\epsilon).
\end{equation}

Proof. We start with the identity 
\[
e^{t\tilde{\mathcal{L}}}|1\rb=e^{t\lambda_{1}}|1\rb+\epsilon\frac{e^{t\tilde{\mathcal{L}}}-e^{t\lambda_{1}}}{\tilde{\mathcal{L}}-\lambda_{1}}|e\rb.
\]
We then obtain $\lb1|e^{t\tilde{\mathcal{L}}}1\rb=e^{t\lambda_{1}}+\epsilon\eta$,
and taking the modulus $\left|\lb1|e^{t\tilde{\mathcal{L}}}1\rb\right|=\left|e^{t\lambda_{1}}\right|+\epsilon\eta'+O(\epsilon^{2})$
with $\left|\eta'\right|\le\left|\lb1|\frac{e^{t\tilde{\mathcal{L}}}-e^{t\lambda_{1}}}{\tilde{\mathcal{L}}-\lambda_{1}}e\rb\right|=O(1).$
Moreover, assuming that $\tilde{\mathcal{L}}$ can be diagonalized,
$\left|\eta'\right|$ does not blows up with $t$, rather 
\begin{align}
\left|\eta'\right| & \le\left\Vert e\right\Vert \left\Vert (\tilde{\mathcal{L}}-\lambda_{1})^{-1}\right\Vert \left(\sum_{j}\left|e^{t\lambda_{j}}\right|\left\Vert P_{j}\right\Vert +\left|e^{t\lambda_{1}}\right|\right)\nonumber \\
 & =(c+1)\left\Vert e\right\Vert \left\Vert (\tilde{\mathcal{L}}-\lambda_{1})^{-1}\right\Vert ,
\end{align}
having set $c=\sum_{j}\left\Vert P_{j}\right\Vert $, since $\mathrm{Re}(\lambda_{j})\le0$
and $t\ge0$.

The same conclusion holds, not surprisingly, using a slightly relaxed
assumption $P_{1}=|1\rb\lb1|+\epsilon X$. Using the normalization
of the projectors $P_{i}P_{j}=\delta_{i,j}P_{j}$ one obtains $\lb1|P_{j}|1\rb=\Tr(P_{j}|1\rb\lb1|)=\Tr[P_{j}(P_{1}-\epsilon X)]$.
The latter expression equals 1 for $j=1$ and $O(\epsilon)$ otherwise.
Hence 
\begin{align}
\lb1|e^{t\tilde{\mathcal{L}}}1\rb & =\sum_{j}e^{t\lambda_{j}}\Tr(P_{j}|1\rb\lb1|)\\
 & =e^{t\lambda_{1}}-\epsilon\Big[\Tr(P_{1}X)+\sum_{j\neq1}e^{t\lambda_{j}}\Tr(P_{j}X)\Big],
\end{align}
and the result holds with $\left|\eta'\right|\le\sum_{j}\left|\Tr(P_{j}X)\right|$.
As can be seen from the absence of the resolvent term, in this case
the error can be made significantly smaller.

Note that if $\text{Re(\ensuremath{\lambda)=0}}$ the leading term
of the coherence does not decay. This fact will be important when
discussing dark or quasi-dark states in topological models.

To gain further insight we analyze the weak and strong dissipative
limits.

\textbf{Fact 2} \emph{(strong dissipative limit). }Assume a linear
geometry and a hopping to dissipation ratio $\left|J_{1}/\Gamma_{2}\right|=\epsilon$
sufficiently small. Then conditions i') and ii') are satisfied and
in particular $\mathcal{C}(t)=e^{-t/\tau_{\mathrm{coh}}}+O\left(\epsilon\right)$
with $\tau_{\mathrm{coh}}^{-1}=2J_{1}^{2}/\Gamma_{2}+J_{1}O(\epsilon^{2})$.

Proof. We consider the off-diagonal terms of Eq.~\eqref{eqn: H}
as a perturbation. The spectrum of the unperturbed system is $\left\{ 0,-\Gamma_{i}/2,i=2,\ldots,N\right\} $,
and the zero eigenvalue has eigenprojector $|1\rb\lb1|$. Using (non-Hermitian)
perturbation theory, the first correction to the zero eigenvalue occurs
at second order and is given by $\lambda_{1}^{(2)}=-2J_{1}^{2}/\Gamma_{2}$.
The corresponding eigenprojector is given by $P_{1}=|1\rb\lb1|+O\left(\epsilon\right)$
so that we are in the condition for fact 1 and the result follows.
Note that, since eigenvalues are continuous in their parameters, as
long as there is no level crossing, $\lambda_{1}$ is the eigenvalue
with real part closest to zero. In other words the coherence time
is given by the \emph{slowest} time-scale of $\tilde{\mathcal{L}}$.

We define $\mathsf{H}=\mathsf{H}_{0}+\mathsf{D}$ where $\mathsf{H}_{0}$
($\mathsf{D}$) is the Hermitian (anti-Hermitian) part of $\mathsf{H}$.
Note that the matrix $\mathsf{D}$ is diagonal in the ``position''
basis $|j\rb$. Since $\mathsf{H}_{0}$ is Hermitian it can be written
as $\mathsf{H}_{0}=\sum_{k}\epsilon_{k}^{(0)}|k\rb\lb k|$, where
$|k\rb$ are the unperturbed eigenvectors. Up to first order, the
eigenvalues of $\tilde{\mathcal{L}}$ become 
\begin{align}
\lambda_{k} & =-i\epsilon_{k}^{(0)}-i\lb k|\mathsf{D}|k\rb\\
 & =-i\epsilon_{k}^{(0)}-\frac{1}{2}\sum_{j=2}^{N}\Gamma_{j}\left|\lb k|j\rb\right|^{2}.\label{eq:small_coupling}
\end{align}
In the isotropic case where all the cavities are equal $\Gamma_{j}=\Gamma$
and the above becomes 
\[
\lambda_{k}=-i\epsilon_{k}^{(0)}-\frac{\Gamma}{2}\left(1-\left|\lb k|1\rb\right|^{2}\right).
\]

Moreover, assume now that the Hamiltonian $\mathsf{H}_{0}$ has a
state localized at the first site: $\exists k_{0}:\,|k_{0}\rb\approx|1\rb$.
This means that $|1\rb\lb1|$ is an approximate eigenprojector of
$\mathsf{H}_{0}$: $P_{k_{0}}\equiv|k_{0}\rb\lb k_{0}|=|1\rb\lb1|+\epsilon Y$
with a small $\epsilon$. Surprisingly $|1\rb\lb1|$ is also an eigenprojector
of $\mathsf{H}$ up to the same order. In fact the first correction
to the eigenprojectors of $\mathsf{H}$ is $P^{(1)}=-P_{k_{0}}\mathsf{D}S-S\mathsf{D}P_{k_{0}}$
where $S$ is the reduce resolvent \cite{kato_perturbation_1995}.
Plugging in $\mathsf{D}=-i(\Gamma/2)\left(\1-|1\rb\lb1|\right)$ we
obtain $P^{(1)}=i(\Gamma/2)\epsilon\left[Y(\left(\1-|1\rb\lb1|\right)S+S\left(\1-|1\rb\lb1|\right)Y\right]$.
Then $|1\rb\lb1|$ is a projector of $\tilde{\mathcal{L}}$ up to
an error $O(\Gamma\epsilon)$. In other words we have the following

\textbf{Fact 3} \emph{(weak dissipative limit). }Assume that the Hamiltonian
$\mathsf{H}_{0}$ has a state localized at the first site: $P_{k_{0}}=|1\rb\lb1|+\epsilon'Y$
with a small $\epsilon'$. For small $\Gamma$ both hypothesis i')
and ii') hold. Moreover Fact 1 holds with $\epsilon=\Gamma\epsilon'$:
$\mathcal{C}(t)=e^{-t/\tau_{\mathrm{coh}}}+O(\epsilon'\Gamma).$ The
coherence time is given in this case by $\tau_{\mathrm{coh}}^{-1}=(\Gamma/2)(1-\left|\lb k_{0}|1\rb\right|^{2})+O(\Gamma^{2})$. 
\end{document}